\documentclass[pre,twocolumn,showpacs,groupedaddress,superscriptaddresst]{revtex4-1}
\usepackage{graphicx,amssymb,color}
\usepackage[dvipsnames]{xcolor}
\usepackage[normalem]{ulem}
\usepackage{datetime}

\newcommand{\beq}{\begin{equation}}
\newcommand{\eeq}{\end{equation}}
\newcommand{\beqs} {\begin{displaymath}}
\newcommand{\eeqs} {\end{displaymath}}
\newcommand{\beqa} {\begin{eqnarray}}
\newcommand{\eeqa} {\end{eqnarray}}
\newcommand{\beqas} {\begin{eqnarray*}}
\newcommand{\eeqas} {\end{eqnarray*}}

\begin{document}

\title{Microscopic model for a granular solid-liquid-like phase transition}

\author{S\'ebastien Auma\^itre$^{1,2}$, Nicol\'as Mujica$^3$}

\affiliation{$^1$SPHYNX, Service de Physique de l'Etat Condens\'e, CNRS UMR 3680, CEA Saclay, F-91191 Gif-sur-Yvette Cedex, France}

\affiliation{$^2$Laboratoire de Physique de l'\'Ecole Normale Sup\'erieure de Lyon, CNRS and Universit\'e de Lyon, 46 all\'ee d'Italie, F-69364 Lyon cedex 07, France}

\affiliation{$^3$Departamento de F\'{\i}sica, Facultad de Ciencias F\'{\i}sicas y Matem\'aticas Universidad de Chile,
Avenida Blanco Encalada 2008, Santiago, Chile}


\begin{abstract}

Forced granular matter in confined geometries presents phase transitions and coexistence. Depending on the system and forcing parameters, liquid-vapor and liquid-solid co-existing states are possible. For the solid-liquid coexistence that is observed in quasi-two-dimensional vibrated systems, both first- and second-order transitions have been reported. Experiments show that particles in the solid cluster move collectively, synchronized with the cell's vibration, in a similar way to the collect-and-collide regime observed in granular dampers.
Here, we present a model that proposes a microscopic origin of this granular phase transition and co-existence. Imposing synchronicity, we model the solid cluster as an effective particle of zero restitution coefficient. In addition, we use the mechanical equilibrium between the two phases, with an equation of state validated for hard spheres relating the horizontal velocities in each phase. Balancing energy input and dissipation per unit time we obtain a global power equation, which relates the characteristic vertical and horizontal velocities to the microscopic relevant parameters (geometric and dissipation coefficients) as well as to the vibration amplitude and solid cluster's size. The predictions of the model compare quite well with our experimental results and with the experimental and dynamic simulation results reported elsewhere.
\end{abstract}

\date{\today}
\maketitle

\section{Introduction}
\label{intro}

Granular matter is a prototype of a strongly non-equilibrium system, which can undergo instabilities, pattern formation, phase transitions, coexistence of states, and coarsening \cite{RMP96,RMP06,Baldassarri2015}. A large amount of simple model experiments and simulations have been developed, providing a detailed understanding of basic granular physics, with interest for non-equilibrium statistical mechanics and nonlinear physics. In particular, noncohesive grains that are confined in a vibrated box present liquid-vapor co-existence \cite{Argentina2002,Khain2002,Roeller2011} or liquid-solid state co-existence \cite{Olafsen1998,Prevost2004,Melby2005,Clerc2008,Castillo2012,Castillo2015}, depending on the forcing and dissipative parameters, as well as on the confinement height. Non-equilibrium absorbing phase transitions have also been reported \cite{Neel2014,Maire2024}. 

From a macroscopic point of view, both liquid-vapor and solid-liquid granular phase transitions and the coexistence of these states have been understood as consequences of the effective equation of state, which relates the granular pressure to the filling density and the effective temperature. In fact, in both cases, these systems obey a macroscopic van der Waals equation of state \cite{Argentina2002,Roeller2011,Clerc2008}; Under the appropriate conditions, the compressibility becomes negative and the system is mechanically unstable, the phase separating into a low-density region in mechanical equilibrium with a more dense one. In addition, quite surprisingly, the interface of these noncohesive granular states evidences the existence of an effective surface tension of the order of $1$~mN/m, which was measured in simulations by analyzing the Laplace law for the static liquid-vapor interface \cite{Clewett2012} and experimentally through capillary-like fluctuations at the solid-liquid interface \cite{Luu2013}.

\begin{figure}[t!]
\begin{center}
\includegraphics[width=8.7cm]{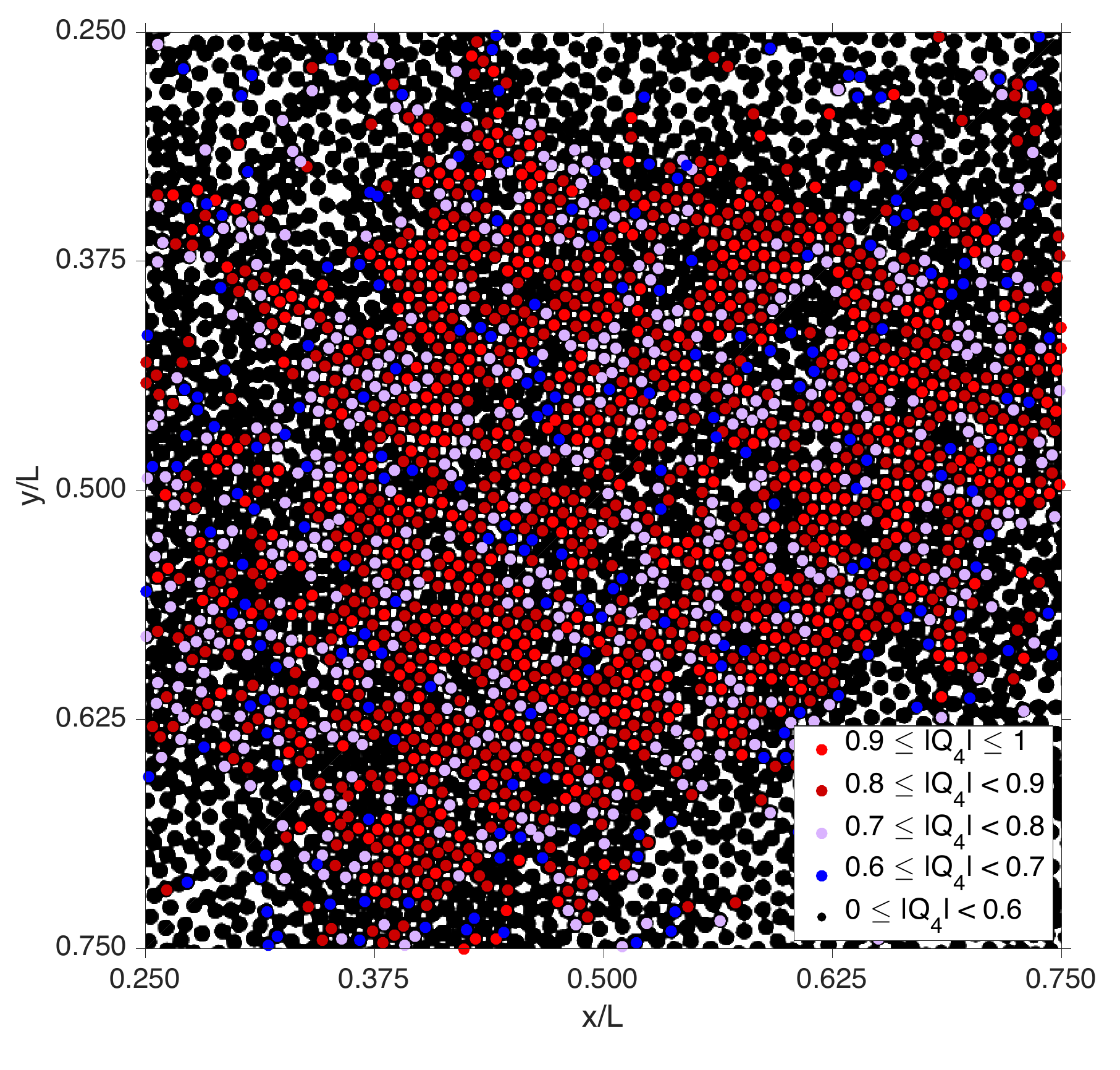}
\caption{Typical image of the solid bi-layered square symmetry crystal in coexistence with the liquid phase ($f = 80$ Hz, $h = 1.94a$, $a = 1$ mm and $A = 0.24a$). The color circles indicate the degree of order, quantified through the absolute value of the four-bond orientational parameter, $|Q_4^j|$. Only the central sector of size  $L/2\times L/2$ is shown.}
\label{fig1}
\end{center}
\end{figure}

Here, we focus on the solid-liquid transition that is observed in quasi-two-dimensional granular systems \cite{Prevost2004,Melby2005,Clerc2008,Castillo2012,Castillo2015}. A set of $N$ noncohesive spheres of diameter $a$ is confined in a shallow box, of horizontal dimensions $L_x$ and $L_y$, and height $L_z \equiv h$. Typically, $L_x = L_y \equiv L \gg a$ and $h \lesssim 2a$, such that the particles only partially overlap. The container is vertically vibrated, sinusoidally. At low driving amplitude the granular system is in an homogeneous fluid state. Counterintuitively, by increasing the forcing amplitude $A$ above a critical amplitude $A_c$, a transition to a more ordered state is observed for a fraction of the particles. Then the system phase separates to solid and liquid states, in mechanical equilibrium (see Fig. \ref{fig1}). The solid cluster is formed by a bilayer of particles with a crystal-like lattice that nearly fills the gap. The lattice structure can be either hexagonal or square, depending on $h/a$ and filling density \cite{Prevost2004,Melby2005}. Although this transition shares similarities with usual phase transitions \cite{Prevost2004,Castillo2012,Castillo2015}, the phase space is quite complex, depending on many parameters, as the driving amplitude and frequency,  confinement height, particle filling fraction, and the restitution coefficients \cite{Prevost2004,Melby2005,Castillo2012,Castillo2015,Vega2008,Lobkovsky2009}. 

\begin{figure*}[t!]
\begin{center}
\includegraphics[width=14cm]{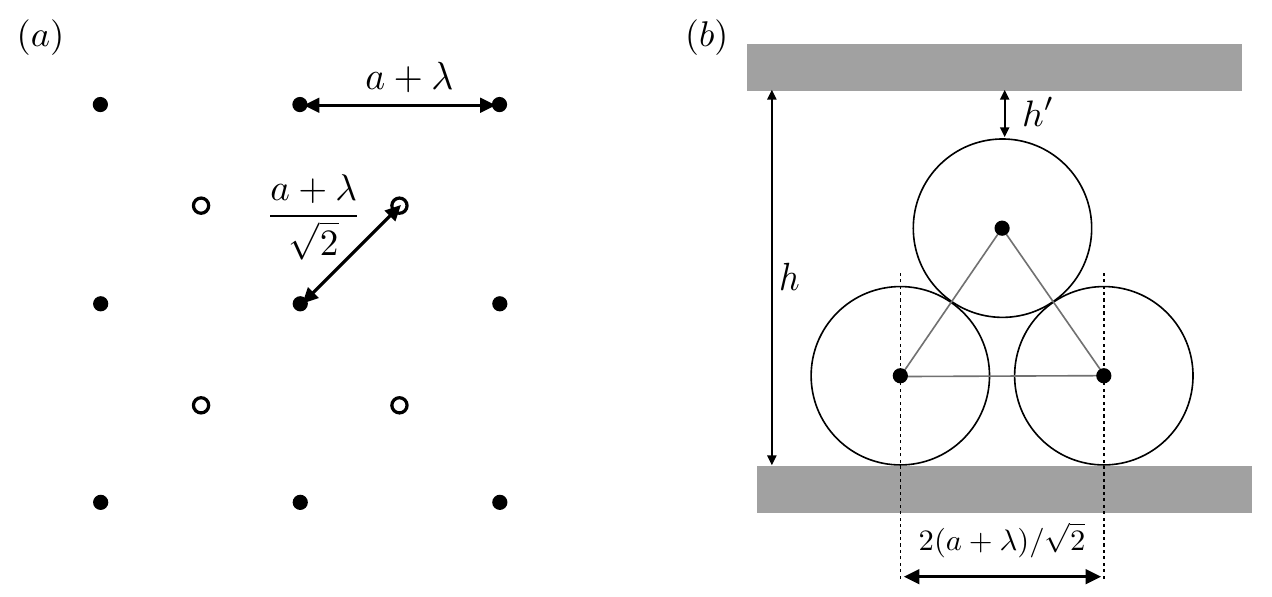}
\caption{(a) Schematic representation of two square interlaced layers for which particles of diameter $a$ are packed with a mean free path $\lambda$. The center of particles in the bottom (top) layer are shown with solid (open) black circles. The 2D projected square lattice has a unit cell length $(a+\lambda)/\sqrt{2}$, its Voronoi area is $\beta = (a+\lambda)^2/2$. (b) Schematic representation of a side view of two particles in the lowest layer supporting another particle on the top layer. The distance $2(a+\lambda)/\sqrt{2}$ is twice the unit cell length of the 2D projected square lattice. }
\label{fig2}
\end{center}
\end{figure*}

In this paper, we present a microscopic model based on experimental observations that explains this transition. Due to internal collisions that efficiently dissipate energy, we assume that the solid phase is in a {\it collect-and-collide} regime, as evidenced in the study of granular dampers \cite{PoschelPRL}. Consequently, as a first approximation, the solid cluster can be considered as single large inelastic particle that is synchronized with the driving, with an effective restitution coefficient $r_{\rm eff} = 0$. Mechanical equilibrium between the two phases and the injected-dissipated power balance are also considered, allowing us to obtain a set of equations relating the system properties, namely the characteristic horizontal and vertical velocities and the solid cluster size. In order to do so, three parameters have to be fixed: the mean free path in the solid phase, the projected 2D disk diameter and the horizontal to vertical velocity anisotropy coefficient in the liquid phase. The values of all these quantities are imposed using experimental data. The model predictions for the critical driving amplitude and cluster size at the onset are compared to the experimental results, showing a good qualitative and semi-quantitative agreement.

\section{Description of the system}
\label{SysDscr}

We consider a set of $N$ hard spheres of diameter $a$ and mass $m$. Collisions between particles are modeled by a restitution coefficient $r'$; between particles and the cell walls it is $r$. Particles are confined to move in a cell of lateral size $L$ in the horizontal direction and of height $h$, with $h <2 a$ in the vertical direction such that the system is quasi-2D. The 2D number density is $\rho_o=N/L^2$ and the 2D projected surface filling fraction is $\phi_o = \rho_o \pi(a/2)^2$.   
The cell is shaken vertically with the forcing $z(t) = A \sin(\omega t)$. Thus, the driving velocity and acceleration are $v(t) =A\omega \cos(\omega t)$ and  $a(t) = - A\omega^2 \sin(\omega t)$, respectively. For all the experimental data presented here $f =\omega/2\pi= 80$~Hz.

At low driving, but with $\Gamma = A \omega^2/g >1$, particles move randomly and collide erratically; the system is in a fluid state. Above a given critical amplitude $A_c$ the system phase separates: $N_l$ particles still move randomly in a liquid state, whereas $N_s = N - N_l$ particles organize themselves into a well ordered cluster, composed of two interlaced square lattices. These are of course fluctuating quantities, with well defined averages for each stationary state, which in turn depend on the forcing parameters. 

The degree of order of those particles in the solid phase is quantified through the four-bond orientational parameter. For each particle, 
\begin{equation}
Q_4^j = \frac{1}{N_{j}} \sum_{s = 1}^{N_{j}} e^{4i\theta{s}^j},
\label{eqn_Q4j}
\end{equation}
where $N_j$ is the number of nearest neighbors of particle $j$ and $\theta_s^j$ is the angle between the neighbor $s$ of particle $j$ and the $x$ axis. For particles in a square lattice, $|Q_4^j | = 1$. The complex phase of $Q_4$ measures  the square lattice orientation respect to the $x$ axis. A typical image of the solid cluster in coexistence with the liquid phase, together with a map of $|Q_4^j|$ for an acceleration $\Gamma= 6.06 > \Gamma_c$, is shown in Fig.~\ref{fig1} ($\Gamma_c = 4.6$ for this configuration). For more details on the transition see Refs. \cite{Clerc2008,Castillo2012,Castillo2015}. 

Particles within the solid cluster have a nonzero horizontal kinetic energy, which implies that they are not in contact all the time. This is evidenced from measurements of their 2D projected Voronoi area, which are larger than the one for particles in contact in this bilayer square lattice \cite{Luu2013}. Then a mean separation between particles that belong to the same layer can be measured, which we will call the mean free path $\lambda$. In Fig. \ref{fig2} we show schematic representations of the top and side views of the 2D projected system, including the definition of $\lambda$. Typically, on average $\lambda \sim 0.1a$. In principle, its value should be determined by the cell's height, crystal structure, and pressure balance between the two phases. However, in the analysis of our model we will assume it is constant, and we will use values obtained from different experimental configurations. We define $u_l$ and $v_l$ as the horizontal and vertical velocities in the liquid state; $u_s$ and $v_s$ are the same velocities in the solid cluster.

Our aim is to understand this solid-liquid transition from a microscopic point of view. In particular we will show that it is possible to determine the fraction of crystallized spheres, $n = N_s/N$, as function of the normalized driving amplitude, $A/a$. This function depends on several parameters, namely the cell's geometry, solid crystal structure, filling density, forcing frequency, restitution coefficients, the horizontal to vertical velocity anisotropy and solid cluster's mean free path. By imposing balance of both mechanical pressure and injected and dissipated energies, we demonstrate that the phase separation occurs at a given critical amplitude $A_c$, and that it can be either continuous or abrupt depending on the system parameters, in good agreement with experimental observations \cite{Castillo2012}. 

\section{Synchronized solid cluster in the collect-and-collide regime}
\subsection{Experimental evidence of the synchronization regime and about the collect-and-collide assumption}
\label{subsec_collect_and_collide}

\begin{figure}[b!]
\centering
\includegraphics[width=8.5cm]{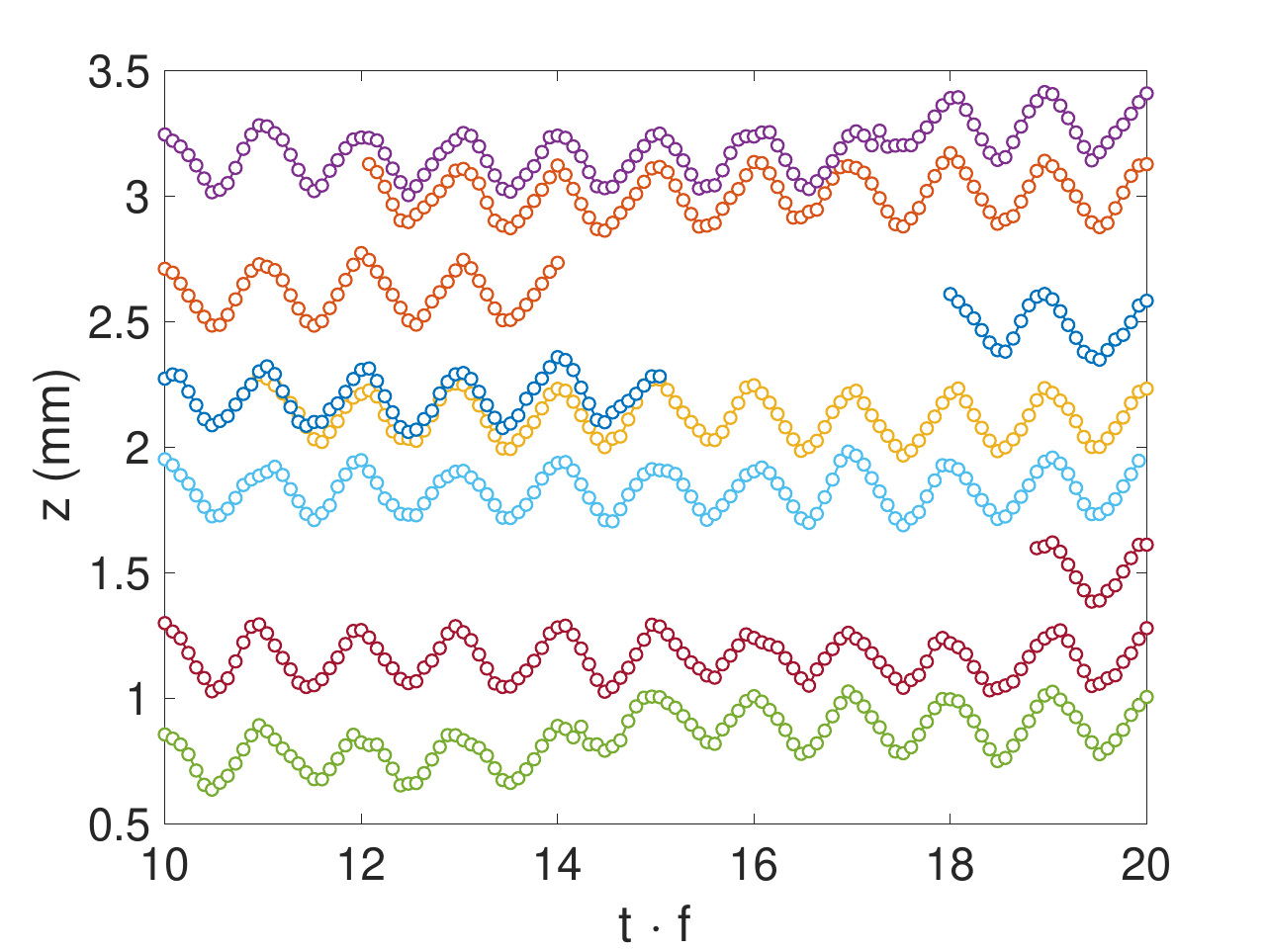}
\caption{Vertical position of 10 particles that have been detected and tracked within the solid phase during 10 oscillation cycles. Data is obtained from images captured from an inclined side view of the experimental cell.}
\label{fig3}
\end{figure}

The synchronized collective regime is obvious from simple observations of the solid cluster visualized laterally. Its periodicity is also easily verified by taking images at a high acquisition rate and by checking the vertical trajectories of particles within the solid cluster.  In Fig.~\ref{fig3} we present experimental measurements of the vertical position of 10 particles detected and tracked in the solid cluster, for exactly 10 periods of oscillation. We obtain such trajectories from a sequence of images acquired with an inclined view of the solid phase; the camera is placed on one side of the experiment with its axis pointing towards the cell's top lid with an angle of about 30 degrees with respect to the horizontal. Most of the trajectories survive the complete time presented in this plot. However, some end and others start in between, which are just consequences of some particles being lost by the tracking algorithm, and either the same or other particles reappearing later during the tracking period. Regardless, the takeaway message from this figure is that particles are indeed moving collectively in the vertical direction, synchronized with the cell's oscillation. In fact, they all perform an up and down movement during one period of oscillation. 

Thus, although it has internal vibration, with horizontal and vertical kinetic energies, we consider the solid cluster as a single particle of mass $N_sm$. In addition, as just shown, the cluster as a whole is synchronized with the forcing vibration. Next, we assume that it is in a {\it collect-and-collide} regime,  introduced previously in the context of granular dampers \cite{PoschelPRL}. In this regime, the cluster behaves as a single perfectly inelastic particle, dissipating all of its kinetic energy at the collision with a wall (either top or bottom lid). More interestingly, in the case of granular dampers, it has been shown that in microgravity conditions the transition from a homogeneous gas state to the condensed collect-and-collide regime occurs when the dissipated energy per cycle is maximized~\cite{PoschelPRL}. In what follows we will review this maximization principle in the zero gravity case, but applied to the solid cluster in our experiment, and we will then generalize it for the non-zero gravity case, which, as we will justify, is the relevant situation for our experiments.

\subsection{Solid cluster as a completely inelastic particle: Zero gravity case}

The solid cluster is composed of $N_s$ particles of mass $m$. We will model this cluster as a completely inelastic particle, such that its restitution coefficient is $r_{\rm eff} = 0$, which is in the collect-and-collide regime described previously.  The cluster is collected in one wall that is moving with velocity $v(t) = A\omega \cos(\omega t)$. At time $t=0$, it will leave the wall when the acceleration changes sign, i.e. at the maximum velocity $A\omega$. Hence, at take-off, the solid cluster velocity is $v_s = A\omega$. Then after a flying time $t_c$, it will collide with the opposite wall, which has a velocity $v_w\equiv v(t_c) =A\omega \cos(\omega t_c)$. The collision time $t_c$ is given by  
\begin{equation}
A\omega\, t_c = A \sin(\omega\cdot t_c) +h'.
\label{impl}
\end{equation}
Here, we consider the free flight distance $h'$, which is related to the mean free path in the cluster and to the cell height by $h'=h-(a+\sqrt{a^2-\beta})$, with $\beta=(a+\lambda)^2/2$ the Voronoi area of a square two-layered square lattice with a mean free path $\lambda$ (see Fig. \ref{fig2} (b)). Equation (\ref{impl}) can be solved numerically.  At every oscillation cycle there are two collisions between the solid cluster and the walls. Then, the kinetic energy that is dissipated per period of the completely inelastic particle is
\begin{eqnarray}
\Delta E_{\rm diss} &=&  N_s m \left(v_s-v_w \right)^2 \nonumber \\
&=& N_s m A^2\omega^2\left(1-\cos(\omega t_c) \right)^2.
\label{EDiss1}
\end{eqnarray}
This quantity is clearly maximum at $\omega t_c=\pi$, which corresponds to perfect synchronicity. In this case, the vibration amplitude is $A=h'/\pi$. In fact, in Ref. \cite{PoschelPRL} it was shown that the transition from a gas state to the condensed collect-and-collide regime occurs precisely at $A = A_ c =h'/\pi$. 

To obtain an explicit expression of the dissipated energy as a function of the driving amplitude $A$, Eqn.~(\ref{impl}) is expanded around the synchronicity condition $\omega t_c=\pi+\epsilon$, with $\epsilon\ll1$. We get $\epsilon\approx-\pi/2+h'/(2A)$ and thus $\omega t_c\approx \pi/2+h'/(2A)$.  Eqn. (\ref{EDiss1}) is then approximated by
\begin{equation}
\Delta E_{\rm diss} \approx N_s m A^2\omega^2\left(1+\sin(h'/2A)\right)^2.
\label{Diss2}
\end{equation}
The measured dissipated energy was shown to be well described by this expression for granular dampers in microgravity \cite{PoschelPRL}.

\subsection{Solid cluster as a completely inelastic particle: Role of gravity}

In the collect-and-collide regime the role of gravity remains important even at high frequency. The liquid-solid transition occurs at a critical acceleration $\Gamma_c$ that is not much larger than $g$; depending on the experimental configuration it is in the range $2-5$. Indeed, whatever is the forcing frequency, the collapsed quasi-particle has to overcome  gravity to leave the wall. 
Considering the experimental observations described in Section \ref{subsec_collect_and_collide}, we are interested in the cases where the quasi-particle collides twice per oscillation period, consecutively with the top and bottom plates. We can no longer consider both collisions per cycle to be equal. 

Thus, the strategy is to compute numerically the takeoff and collision times of an effective completely inelastic particle ($r_{\rm eff}=0$) under gravity that is confined between two horizontal plates separated by a distance $h'$, which oscillate sinusoidally with amplitude $A$ and frequency $f$. Then, once the takeoff and collision times are determined, the computation of the collision velocities is straight forward. The effective particle is the solid cluster with square crystalline structure, composed of $N_s$ particles of mass $m$, with an average Voronoi area $\beta$. Once the latter is fixed, also the mean free path $\lambda$ and the free flying distance, which is now an effective height, $h'$, become fixed. 

\begin{figure}[t!]
\centering
\includegraphics[width=8.5cm]{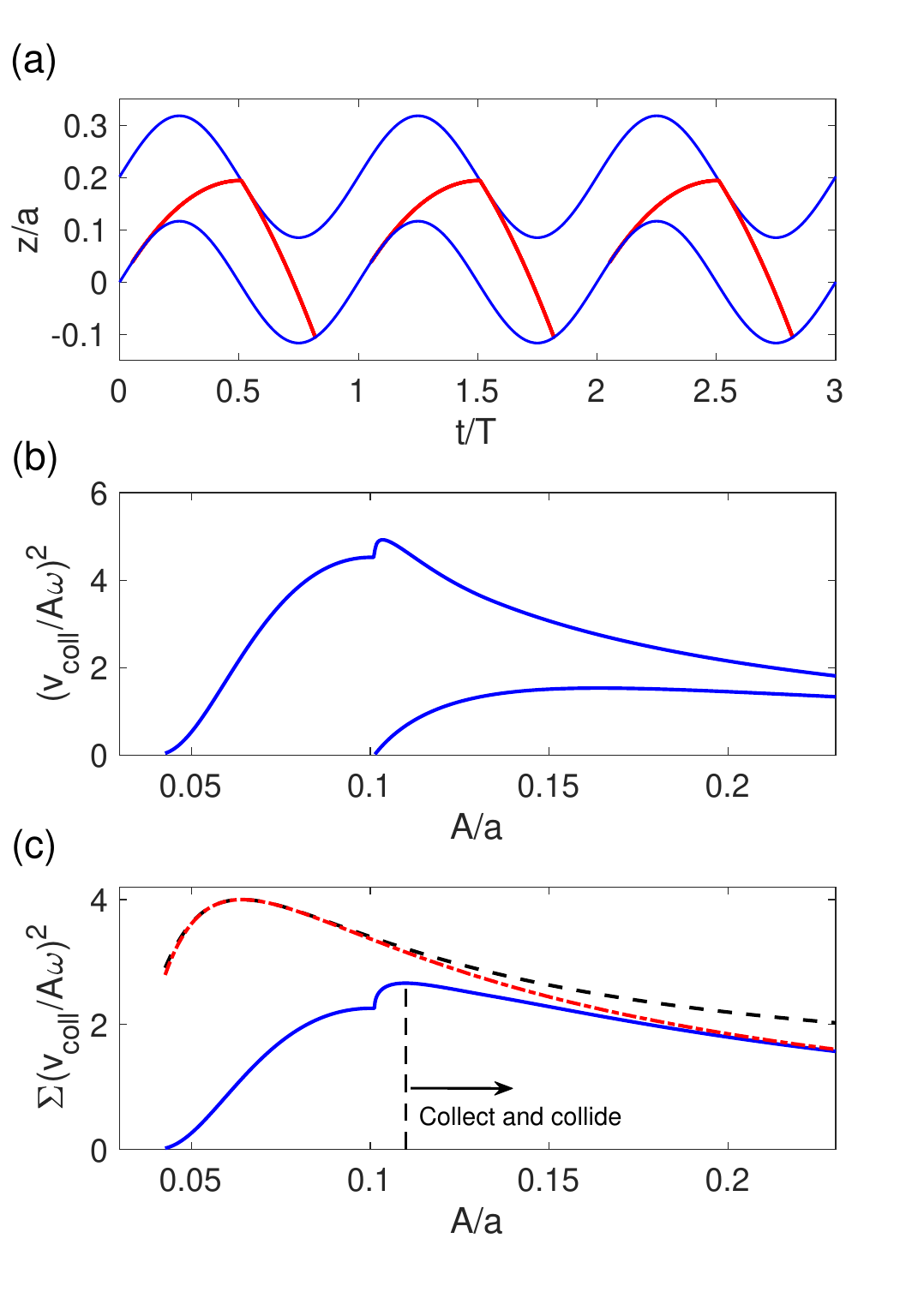}
\caption{(a) Example of quasi-particle trajectories computed numerically, with $r_{\rm eff}~=~0$, $\Gamma = 3$, $f = 80$ Hz, $\lambda/a~=~0.1$ and $h/a = 1.83$ ($A/a = 0.1165$ and $h'/a=0.2015$). From this kind of trajectory, the collision velocity $v_{\rm coll} = v_s - v_w$ can be computed at each impact. (b) Normalized $v_{\rm coll}^2$ as function of $A/a$, for the same set of parameters of figure (a).  For $A<A^* = 0.1013a$ ($\Gamma^* = 2.61$) only one collision per period occurs. (c) Normalized dissipated energy per oscillation cycle, $\Delta E_{\rm diss}/(mN_s A^2\omega^2)$ (continuous line) as function of $A/a$ for the same parameters as in (b). The dissipation is maximum at $A_{\rm max} =  0.1099a$ ($\Gamma_{\rm max} = 2.83$), indicated by the vertical dashed line. The dashed-dotted line (dashed line) shows as comparison the same quantity in absence of gravity, given by the exact Eqn.~(\ref{EDiss1}) (approximated Eqn.~(\ref{Diss2})).}
\label{fig4}
\end{figure}

In Fig. \ref{fig4}a we present an example of a periodic trajectory of the completely inelastic particle under gravity that is confined between two oscillating walls. The forcing, geometrical and quasi-particle parameters are specified in the figure caption. The effect of gravity is evident at such vibration amplitude ($\Gamma = 3$, $f = 80$ Hz), as it is clear that both the upward and downward trajectories are parabolic. In Fig. \ref{fig4}b we plot $v_{\rm coll}^2$ normalized by $(A\omega)^2$ versus the dimensionless amplitude $A/a$. Here, $v_{\rm coll} = v_s-v_w$ is the collisional velocity. Below $A^*= 0.1013a$ only one collision per oscillation period occurs, which indicates that the particle does not reach the top wall during its free flight, colliding always with the bottom wall. Above this amplitude, two collisions per period occur, with clearly two distinct collisional velocities, which tend to equalize for larger forcing.

In the case of two collisions per oscillation, the dissipated energy per period is 
\begin{equation}
\Delta E_{\rm diss}=\frac{1}{2} N_sm \left(v_{\rm coll,1}^2+ v_{\rm coll,2}^2\right),
\label{EDiss2}
\end{equation}
where $v_{\rm coll,i}$, with $i = \{1,2\}$, correspond to the two distinct collisional velocities at the top and bottom plates respectively. Below $A^*$ only one collision per cycle occurs, at the bottom wall, and the dissipated energy per collision is simply $\Delta E_{\rm diss}= N_s m v_{\rm coll}^2/2$.

In Fig. \ref{fig4}c we present the normalized dissipated energy per oscillation cycle, $\Delta E_{\rm diss}/(mN_s A^2\omega^2)$, as function of $A/a$, using the same set of parameters as before. As in the zero gravity case, shown with the dash-dotted and dashed line curves, this quantity presents a maximum at a given amplitude, $A_{\rm max} = 0.1099a$. It is however significantly larger than the zero gravity prediction $A_c = h'/\pi = 0.0641a$. Nonetheless, the qualitative behavior is quite similar for amplitudes above the maximum of each curve. Indeed, the solution of Eqn.~(\ref{EDiss2}) converges to the exact solution of the zero gravity case for high amplitudes, as expected. Finally, in analogy to the zero gravity case, it what follows we will consider the collect-and-collide regime to be in place for $A \geq A_{\rm max}$. 

\section{Fraction of particles in the solid cluster}
\subsection{Mechanical equilibrium}

Since we have two coexisting phases, they must be in mechanical equilibrium. The pressure exerted by the disordered liquid phase, $P_l$, balances the pressure in the solid ordered phase, $P_s$,  i.e. $P_s=P_l$. The general form of the equation of state has kinetic and collisional terms, given by \cite{Luding2009}
\begin{equation}
P = \frac{m}{2} \frac{N}{S} \langle u^2\rangle \left [ 1 + 2\phi g(\phi) \right ],
\label{eqn1}
\end{equation} 
where the number surface density $N/S$ in each phase is related to the surface fraction $\phi = N S_o/S$, with $S_o$ the quasi 2D surface area (to be defined). In a 2D system composed of disks, $S_o = \pi a^2/4$, with $a$ the particle diameter. The function $g(\phi)$ is the particle correlation function evaluated at contact, which is function of the surface fraction. The collisional term of the equation of state, Eqn.~(\ref{eqn1}), can be written as \cite{Luding2009}
\begin{equation}
\phi g(\phi) = \frac{l_o}{l(\phi)},
\label{eqn3}
\end{equation} 
with $l_o = \sqrt{2\pi}a/8$ for 2D disks (we will assume this value later but it can be also let free as a adjustable parameter for our quasi 2D system of spheres confined between two plates with height $h$).

Thus, the equation of state can be rewritten as 
\begin{equation}
P = \frac{m}{2} \frac{N}{S} \langle u^2\rangle \left [ 1 + 2 \frac{l_o}{l(\phi)} \right ].
\label{eqn4}
\end{equation} 
The solid phase has a well defined structure. Therefore, the surface  occupied by the cluster is directly related to the number of particles, $S_s=\beta N_s$, where $\beta=(a+\lambda)^2/2$ has been introduced previously. Here it represents the inverse of the solid cluster's density for the cubic lattice structure. By measuring each solid particle's Voronoi area, $\beta$, we obtain a measurement of the mean free path $\lambda$ in the solid phase. Thus, for the solid phase the pressure is  
\begin{equation}
P_s = \frac{m}{2} \frac{1}{\beta(\lambda)} \langle u_s^2\rangle \left [ 1 + 2 \frac{l_o}{\lambda} \right ],
\label{eqn5}
\end{equation}
where we have identified the mean free path in this phase by the measured quantity $\lambda$. The prefactor $2$ in the collisional term can be replaced by $1+r$ for inelastic hard spheres, where $r$ is the particle-particle collision restitution coefficient. 

The philosophy here is that in the solid phase, as such high surface fractions, it is more difficult to have a good model to predict the collisional term, so it is difficult to have a good model for $g(\phi)$, so we used measured values of $\lambda$ and $u_s$, with a reasonable estimate for $l_o$. 

On the other hand, for the liquid phase it is not so straight forward to measure the mean free path, and we can use a model valid for moderate densities \cite{Luding2009}. Thus, we can write the equation of state for the liquid phase as 
\begin{equation}
P_l = \frac{m}{2} \frac{N_l}{S_l} \langle u_l^2\rangle \left [ 1 + 2\phi g(\phi) \right ],
\end{equation} 
with $\phi_l = N_l S_o^l/S_l$. Here, the surface $S_o^l$ in the liquid phase turns out to be the only adjustable parameter. Indeed, because the system is quasi 2D, we will treat it as an effective real 2D system (of disks) with an effective diameter $a_{\rm eff}$.

Due to the vertical confinement there is a minimum distance between particle centers, thus there is a minimum effective diameter given by 
\begin{equation}
a_{\rm min} = \sqrt{a^2 - (h-a)^2}.
\end{equation}
We will parametrize the effective diameter as
\begin{equation}
a_{\rm eff}(\kappa) = a_{\rm min} + \kappa (a -  a_{\rm min}), 
\end{equation}
with $0 \leq \kappa \leq 1$. Then, for the effective particle surface we have 
\begin{equation}
S_o^l = \frac{\pi a_{\rm eff}(\kappa)^2}{4}.
\end{equation}
Using $N_l = N - N_s$, $S_l = S - S_s$, $n = N_s/N$ and $\rho_o = N/S$, we get 
\begin{equation}
\frac{N_l}{S_l} = \frac{(1-n)\rho_o}{(1-\beta \rho_o n)}.
\end{equation}
So, the equation of state in the liquid state is 
\begin{equation}
P_l = \frac{m}{2} \frac{(1-n)\rho_o}{(1-\beta \rho_o n)} \langle u_l^2\rangle \left [ 1 + 2\phi(n,\kappa) g(\phi(n,\kappa)) \right ],
\end{equation} 
where we have explicitly introduced the dependence of the liquid surface fraction as function of the solid number fraction $n=N_s/N$ and the effective diameter parameter $\kappa$ (we recall that in the liquid phase $\phi = N_l S_o^l/S_l$). 

The pair correlation function at contact is modeled as function of the surface fraction
\begin{equation}
g(\phi) \equiv g_2(\phi) = \frac{1-7\phi/16}{(1-\phi)^2}. 
\label{eqng}
\end{equation} 
A higher order correction is possible (function $g_4(\phi)$ in \cite{Luding2009}), but for the moderate surface fractions that we have in our experiment this correction is negligible. 

In a stationary state of solid-liquid coexistence, we have the pressure balance $P_s = P_l$, neglecting surface tension effects, which we use to obtain the ratio between the RMS square velocities 
\begin{equation}
\frac{ \langle u_s^2\rangle}{ \langle u_l^2\rangle} = f(n), 
\label{eqn_fn}
\end{equation} 
with 
\begin{equation}
f(n) = \frac{\frac{(1-n)\rho_o}{(1-\beta(\lambda) \rho_o n)} \left [ 1 + 2\phi(n,\kappa) g(\phi(n,\kappa)) \right ]}{\frac{1}{\beta(\lambda)} \left [ 1 + 2 \frac{l_o}{\lambda} \right ]}.
\label{eqn_fn2}
\end{equation}

\subsection{Dissipated power of the solid cluster as a completely inelastic particle}

For the solid cluster, modeled as a completely inelastic particle of mass $N_s m$ and effective restitution coefficient $r_{\rm eff}= 0$, we can express its injected and dissipated energies per oscillation period as 
\begin{eqnarray}
 \Delta E_{\rm inj}  &=& N_s m  \sum{v_w(v_w-v_s)},  \\ \label{PIS}
 \Delta E_{\rm diss}&=&\frac{1}{2} N_s m \sum{ (v_s-v_w)^2},   \label{PDS}
\end{eqnarray}
where $\sum$ represents the addition of the collision events. 
We now consider the dissipated power of this effective particle, which is the dissipated energy per unit time 
\beq
D_s^* =  \Delta E_{\rm diss} \frac{\omega}{2\pi}.
\eeq
We can also express the dissipated power of the solid cluster considering its internal structure, kinetic energy and composition. There should be two dissipation terms: (1) a contribution from vertical collisions of the whole solid cluster with the top and bottom walls as well as internal vertical collisions within the solid cluster, and (2) a contribution due to internal collisions in the horizontal degree of freedom between particles of a same layer. Thus, we obtain
\begin{eqnarray}
\langle D_s \rangle &=& \frac{1}{2} N_s m (1-r^2) \left ( \sum v_{\rm coll}^2 \right ) \nu_s \nonumber \\
&+& \frac{1}{2} N_s m (1-r^2) \langle \Delta u_s^2 \rangle \nu_s'.
\end{eqnarray}
Here, the first term corresponds to vertical particle-wall and interlayer particle-particle collisions. The second to the horizontal particle-particle collisions, of particles in the same layer. For simplicity, we have considered both restitution coefficients (particle-wall and particle-particle) to be equal, $r' = r$. The brackets $\langle\,\, \rangle$ correspond to time averages. The velocity of two colliding particles are independent, thus their relative horizontal velocity obeys $\langle \Delta u_s^2 \rangle = 2 \langle u_s^2\rangle$. 

The solid cluster collides twice per period with the top and bottom walls. The two collisions are considered in the definition of $\sum v_{\rm coll}^2 = \sum (v_s-v_w)^2$,  so we expect the collisional frequency $\nu_s$ to scale as $\omega/2\pi$. However, the solid cluster is composed by two interlaced square layers. It is not completely compact as evidenced by the observation of a non-zero horizontal mean free path $\lambda$. Such dilation should also be present in the vertical direction. Thus, at each collision of the solid cluster with the top or bottom wall, a number of internal collisions within the bi-layer should occur. Simply put, a particle of one layer will collide vertically a certain number of times with its four neighbors of the other layer. Then, we propose 
\beq
\nu_s = \frac{\omega}{2\pi}\delta,
\eeq
where $\delta$ takes into account the internal vertical collisions. It will be obtained by fitting the experimental data, as will be explained below.

Concerning the horizontal collisions within the solid cluster, the particle-particle collisional frequency is 
\beq
\nu_s' = \frac{\langle | u_s | \rangle}{\lambda} = \frac{\sqrt{\pi} \sqrt{ \langle u_s^2 \rangle}}{2 \lambda},
\eeq
where the average speed and RMS velocities are related by $\langle | u_s | \rangle = \sqrt{\pi}\sqrt{\langle  u_s ^2\rangle}/2$ for Maxwellian distributions.

The last step is to consider that the dissipated power of the effective completely inelastic solid cluster that collides twice per period with the top and bottom walls must equalize the one of the bi-layer with both internal and wall collisions, namely $D_s^* = \langle D_s \rangle$. This allows to obtain an expression for the average horizontal velocity of the solid cluster
\beq 
\frac{\sqrt{\langle u_s^2\rangle}}{A\omega} = \left [ \frac{\lambda [1-(1-r^2)\delta ]\sum (v_{\rm coll}/A\omega)^2}{2 A\pi^{3/2}(1-r^2)}\right ]^{1/3}.
\label{ecn_us}
\eeq 
This expression will be compared with experimental data in section \ref{sec_experiments}. The fit of $\sqrt{\langle u_s^2\rangle}/A\omega$ as function of $A$ and $\lambda$ will allow us to obtain estimates for $\delta$.

\subsection{Injected and dissipated power balance}

Until now we have computed the dissipated power of the effective particle
\beq 
D_s^* =  \Delta E_{\rm diss} \frac{\omega}{2\pi} = \frac{1}{2} N_s m \sum{ (v_s-v_w)^2} \frac{\omega}{2\pi}.
\eeq
The same can be done for the injected power
\beq
I_s^* = \Delta E_{\rm inj} \frac{\omega}{2\pi} = N_s m  \sum{v_w(v_w-v_s)} \frac{\omega}{2\pi}.
\eeq
We consider now the complete power balance equation
\beq
\frac{dE}{dt} = I_s^* + \langle I_l\rangle - D_s^* - \langle D_l\rangle = 0,
\eeq 
where 
\begin{eqnarray}
\langle I_l\rangle&=& N_l m (1+r) \langle v_w (v_w-v_l)\rangle \nu_l,\\
\label{PIG}
\langle D_l\rangle&=&\frac {1}{2} N_l m(1-r^2)\langle (v_w-v_l)^2\rangle \nu_l \nonumber \\ 
&+&\frac {1}{2} N_l m(1-r^2)\langle \Delta u_l^{2}\rangle \nu_l',
\label{PDG}
\end{eqnarray}
are the injected and dissipated powers in the liquid phase. As before, the relative horizontal velocity in the liquid phase obeys $\langle \Delta u_l^{2}\rangle = 2 \langle u_l^{2}\rangle$. Now, the particle-wall and particle-particle collisional frequencies in the liquid phase are  
\begin{eqnarray}
\nu_l &=& \frac{\langle | v_l | \rangle}{h-a} = \frac{\sqrt{\pi} \sqrt{ \langle v_l^2 \rangle}}{2 (h-a)},\\
\nu_l' &=& \frac{\langle | u_l | \rangle}{l(\phi)} = \frac{\sqrt{\pi} \sqrt{ \langle u_l^2 \rangle}}{2 l(\phi)},
\end{eqnarray}
where $l(\phi)$ is given by definitions (\ref{eqn3}) and (\ref{eqng}). Unlike the solid phase, the particle and wall velocities of the liquid phase are not correlated, therefore $\langle v_w v_l \rangle = 0$. The average squared wall velocity is $\langle v_w^2 \rangle = A^2\omega^2/2$. Moreover, we relate the vertical and horizontal velocities in the liquid phase by an anisotropy parameter $\alpha$, thus $\sqrt{\langle v_l^2 \rangle} = \alpha \sqrt{\langle u_l^2 \rangle}$.

Finally, putting all together into the power balance equation, we obtain the following equation for fraction of particles in the solid phase as a function of the driving amplitude, frequency, and all the other system parameters,
\begin{widetext}
\begin{equation}
G(n,A) = \frac{n (h-a)}{\pi^{3/2}A}\left [ 2\sum \tilde{v}_w \tilde{v}_{\rm coll} + \sum \tilde{v}_{\rm coll}^2\right ]
+ \frac{(1-n) \alpha \langle \tilde{u}_s^2 \rangle^{1/2}}{f(n)^{1/2}}\left [ \frac{\langle \tilde{u}_s^2\rangle (1-r^2)}{f(n)}\left ( \alpha^2+  \frac{2(h-a)}{\alpha l(\phi)}\right )-\frac{(1+ r)^2}{2} \right ]=0,
\label{eqnGnA}
\end{equation}   
\end{widetext}
where $\tilde{v}_w = v_w/A\omega$, $\tilde{v}_{\rm coll} = v_{\rm coll}/A\omega$ and $\tilde{u}_s = u_s/A\omega$ are the normalized velocities.

\section{Experimental procedures and results}
\label{sec_experiments}

\subsection{Experimental procedures and configurations}

In order to check different aspects of the phenomenological model introduced before, we now present and diskuss some experimental results. We have previously shown that the solid-liquid transition can be of first or second order, depending on experimental parameters. Indeed, in Ref. \cite{Castillo2012}, by studying the global average of the four-fold global orientation parameter $\langle |Q_4|\rangle$, with $Q_4^j$ defined by Eqn. (\ref{eqn_Q4j}), we demonstrate that for lower density $\rho_o$ and cell height $h$ the transition is of the first order, while for larger $\rho_o$ and $h$, it is of the second order. These experiments were performed for different but fixed $\rho_o$ and $h$, and a constant forcing frequency $f = 80$ Hz. The transition was reached by slowly increasing the vibration amplitude~$A$. 

\begin{figure}[t!]
\centering
\includegraphics[width=8cm]{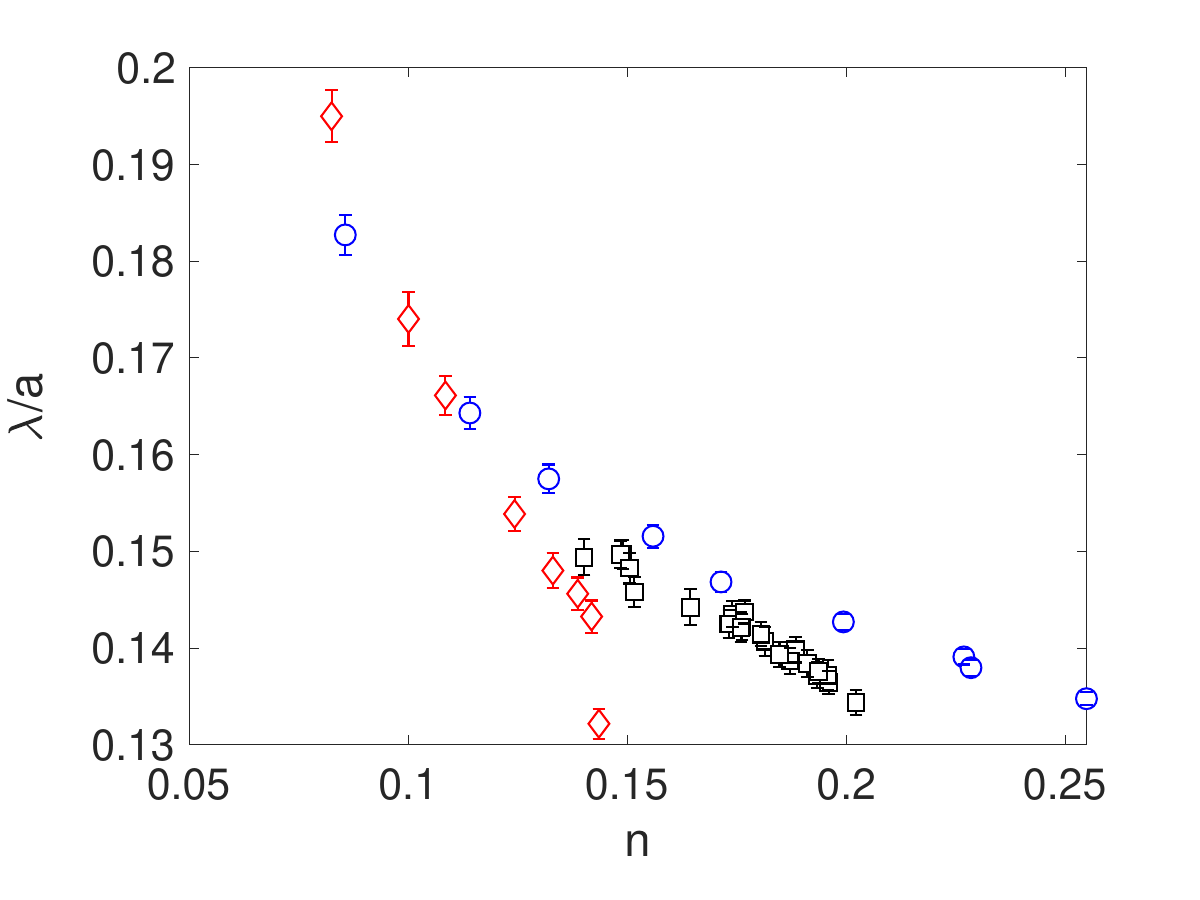}
\caption{Average normalized mean free path $\lambda/a$ as function of the solid phase fraction $n = N_s/N$. Error bars are computed form the standard deviation. Data for the three configurations are presented: C1~($\diamond$), C2a ($\circ$) and C2b ($\square$). Data is presented for $\Gamma>\Gamma_c$.}
\label{fig5}
\end{figure}

Here, we present three new sets of results, which we name experimental configurations C1, C2a and C2b. The experimental setup and procedures are the same as in \cite{Castillo2012,Castillo2015}. 
For all these new experimental runs, the bottom and top glass ITO coatings, used to prevent electrostatic charging, correspond to the thick coating presented in \cite{Castillo2015} (experiment type B in this reference). Particles are stainless steel spheres, $a = 1$ mm diameter.

The first configuration, C1, has the same height as the one presented in Ref. \cite{Castillo2012}, $h = (1.83\pm 0.02)a$, but with $N = 9757$ particles, thus the 2D projected surface filling fraction is $\phi_o = \pi N a^2/(4 L^2) = 0.7663$. The transition is of first order, with $\Gamma_c = 2.0$, and  $\Gamma$ is varied between $2.07$ and $2.88$.

Concerning the second order type transition, configuration C2a also has the same height as in Ref.~\cite{Castillo2012}, $h = (1.94\pm0.02)a$, but now $\Gamma = 6.65\pm0.01$ is kept constant, well above $\Gamma_c$ (the critical acceleration varies with density, but is $\sim 4.5$). In this case, by changing $N$, we vary the surface fraction of particles $\phi_o$, specifically between $0.851$ and $0.919$. On the contrary, configuration C2b is done also with $h = (1.94\pm0.02)a$ but at constant $N = 11773$, implying constant $\phi_o = 0.925$. The normalized acceleration is varied between $2$ and $6$, and the critical acceleration is $\Gamma_c = 4.6$.

These three configurations are studied in detail below, comparing measurements with our model predictions. These sets of experimental results complement each other and allow us to demonstrate the generality of the results and the predictability of the model that will be diskussed. Table \ref{table1} summarizes the experimental configuration parameters.

Two important quantities that are used in these comparisons are the solid phase particle's mean free path $\lambda$ and the system's filling density $\rho_o$ (alternatively the solid particle's Voronoi area $\beta$ and $\rho_o$). In Fig. \ref{fig5} we present the average $\lambda/a$ for each configuration as function of the measured solid phase fraction $n = N_s/N$. All the data correspond to situations where the system has phase separated into a solid cluster coexisting with the liquid phase. In each case, the most probable value (the mode) of the measured quantities is relatively constant, ${\rm Mo}(\lambda)\approx 0.08a$ for C1 and ${\rm Mo}(\lambda)\approx 0.12a$ for C2.
Quite surprisingly, the average $\lambda$ is relatively independent of the configuration at low $n$; in general it decreases as $n$ increases, with a slight configuration dependence for $n\gtrsim 0.15$. The global average values with their corresponding standard deviations are $\langle \lambda/a \rangle= 0.157\pm 0.020$, $0.151\pm0.015$ and $0.142\pm0.005$. These averages are obtained in time first, which corresponds to each data point in Fig. \ref{fig5}, and then averaged for all experimental runs for each configuration (that is for all $A$ for C1 and C2b and for all $\phi_o$ for C2a).  

As stated before, $\rho_o a^2$ is constant for both C1 and C2b, because the number of particles $N$ is fixed. Small variations are measured due to small errors in the detection of particles, which at maximum is about $\pm10$ particles per image. For C2a, where we keep $\Gamma$ constant and vary $N$, $\rho_o a^2$ increases linearly with $n$, as expected.  The global averages are $\langle \rho_o a^2 \rangle= 0.9757\pm0.0001$, $1.127\pm0.028$ and $1.178\pm0.001$ for C1, C2a and C2b, respectively. 

\begin{table}[t!]
\caption{Summary of experimental parameters for the three studied configurations. Forcing frequency and particle diameter are fixed, $f=80$ Hz and $a = 1$ mm.}
\begin{center}
\begin{tabular}{|c|c|c|c|}
\hline
  	& C1 			& C2a 			& C2b \\
\hline
\hline
$\,\,h/a\,\,$	& $1.83\pm 0.02$ 	&  $1.94\pm 0.02$	& $1.94\pm 0.02$ \\
\hline
$\,\,\phi_o\,\,$	& $0.766$ 	&  $0.851 - 0.919$	& $0.925$ \\
\hline
$\,\,\Gamma \,\,$	& $2.07-2.88$ 	&  $6.65$	& $2.01 - 6.06$ \\
\hline
$\,\,A/a \,\,$	& $0.080-0.112$ 	&  $0.258$	& $0.078-0.235$ \\
\hline
$\,\,\langle \rho_o a^2 \rangle \,\,$	& $0.9757\pm0.0001$ 	&  $1.127\pm0.028$	& $1.178\pm 0.001$ \\  
\hline
$\,\,\langle \lambda/a \rangle \,\,$	& $0.157\pm0.020$ 	&  $0.151\pm0.015$	& $0.142\pm 0.005$ \\  
\hline
\end{tabular}
\end{center}
\label{table1}
\end{table}%

\subsection{Experimental results: Equation of State}

\begin{figure}[t!]
\centering
\includegraphics[width=8cm]{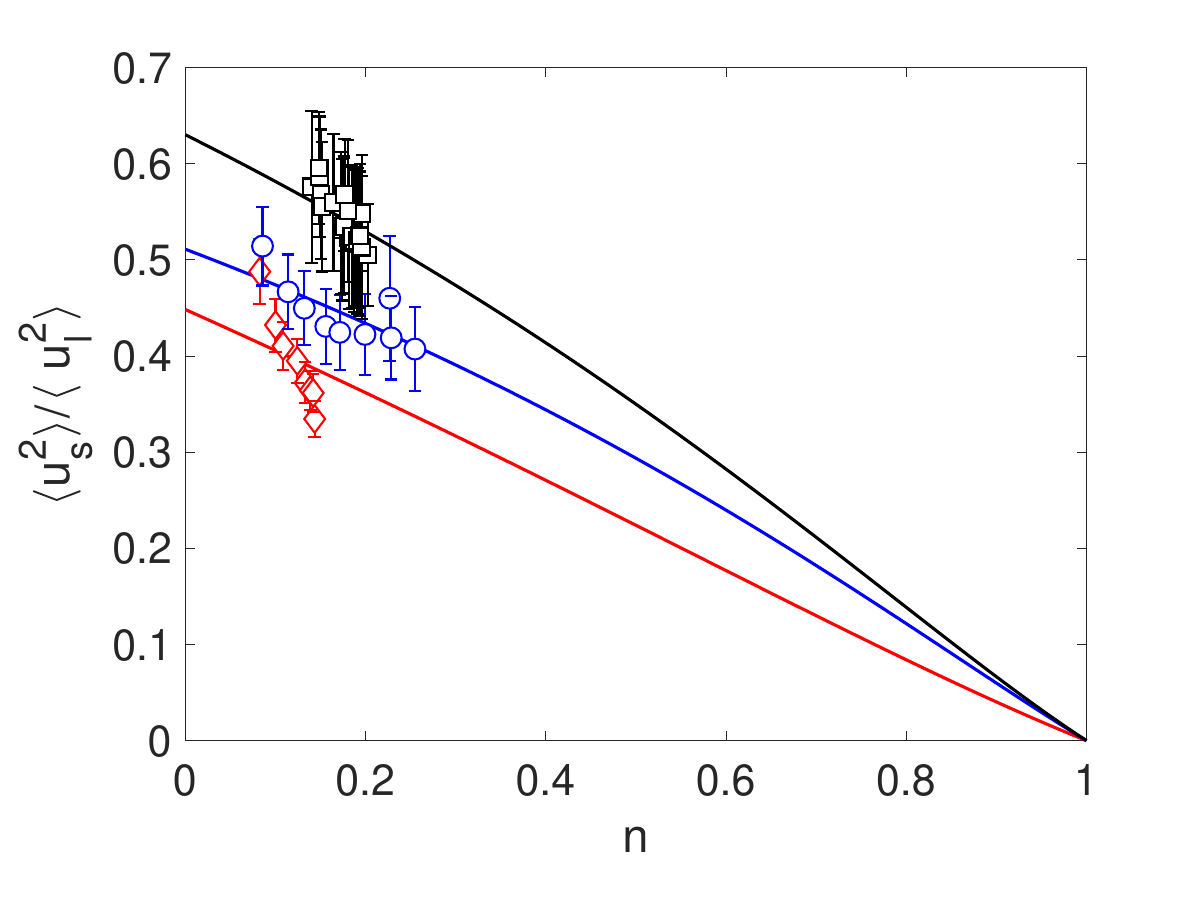}
\caption{Ratio of the average horizontal kinetic energies $\langle u_s^2\rangle/\langle u_l^2\rangle$ as a function of the fraction of particles in the ordered phase, $n$, for the experiments C1 ($\diamond$), C2a ($\circ$) and C2b ($\square$). Solid lines correspond to the fits of Eqn. (\ref{eqn_fn}), using the measured average values of $h$, $\beta$, $\lambda$ and $\rho_o$, and with only one adjustable parameter, the effective diameter parametrization constant $\kappa$. We obtain $\kappa  = 0.476\pm 0.046$ for C1, $\kappa = 0.577\pm 0.013$ for C2a and $\kappa = 0.618\pm 0.004$ for C2b.}
\label{fig6}
\end{figure}

The mechanical equilibrium between phases holds beyond the threshold. It can be checked experimentally at all forcing amplitudes where the two phases coexist. In Fig. \ref{fig6} we present the ratio ${ \langle u_s^2\rangle}/{ \langle u_l^2\rangle} $ as function of the solid number fraction $n$, obtained for the three experimental configurations. The solid lines correspond to the theoretical prediction of equations (\ref{eqn_fn}) and (\ref{eqn_fn2}) using average values for $\beta$, $\lambda$, $\rho_o$ and $h$. The only fitting parameter is $\kappa$, for which we obtain $\kappa  = 0.476\pm 0.046$ for C1, $\kappa = 0.577\pm 0.013$ for C2a and $\kappa = 0.618\pm 0.004$ for C2b. The errors are given by the $95\%$ confidence range of the adjusted parameter. The length $l_o$ is fixed to $\sqrt{2\pi}a/8$ as for 2D disks. Despite the scatter in the data, and the variations of $\lambda$ (thus of $\beta$) and $\rho_o$, the agreement between the model and experiment is surprising good considering the simplicity of the equation of state, at least for the range of $n$ that has been covered. 

\begin{figure}[t!]
\centering
\includegraphics[width=8cm]{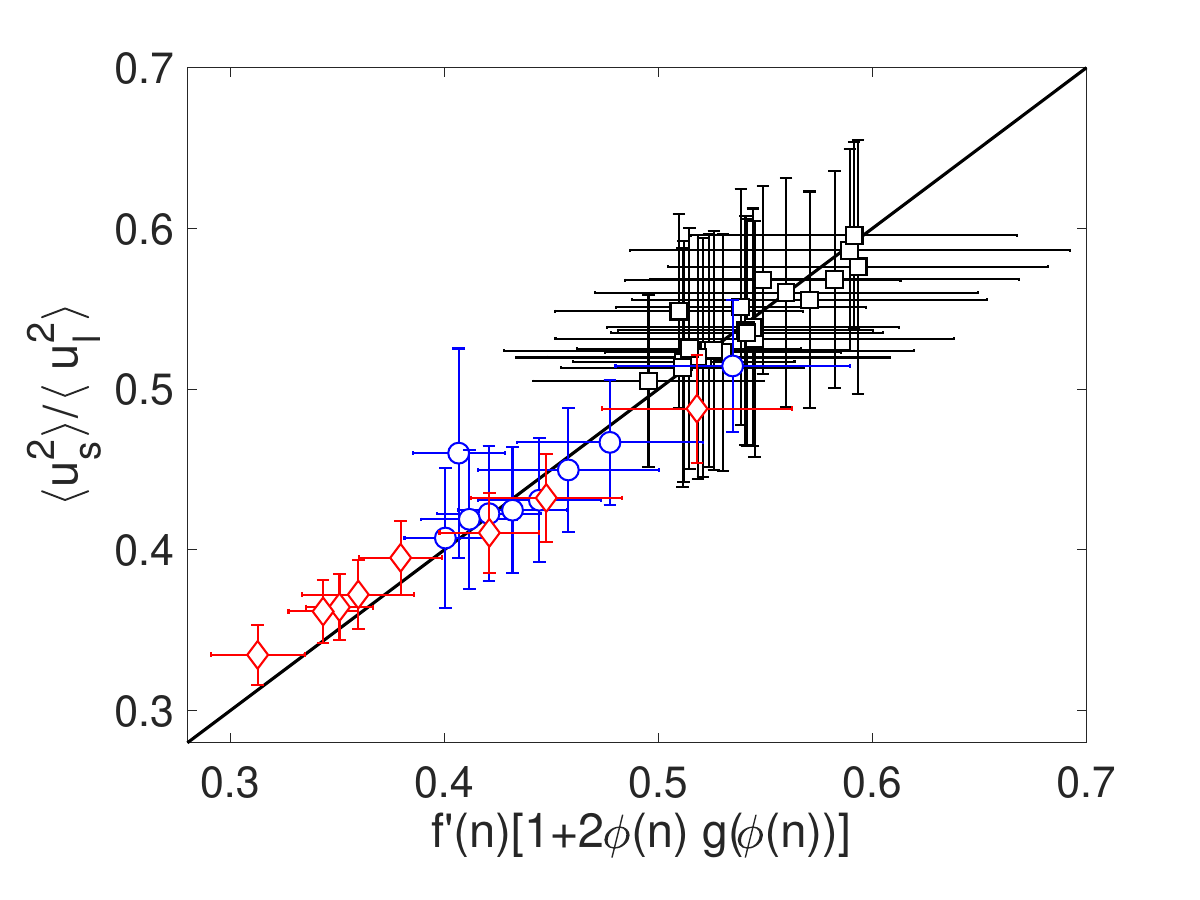}
\caption{Ratio of the average horizontal kinetic energies $\langle u_s^2\rangle/\langle u_l^2\rangle$ versus $f'(n) [1+2\phi(n)g(\phi(n))]$. Results for the three configurations are presented: C1~($\diamond$), C2a~($\circ$) and C2b ($\square$). This is a better verification of the validity of the equation of state as the parameters $\beta$, $\lambda$ and $\rho_o$ are no longer considered fixed by their global averages; instead, for each set of data (that is for each $A$ for C1 and C2b or each $\phi_o$ for C2a) we considerer their time average measured values. }
\label{fig7}
\end{figure}

A better check of the equation of state can be done by recognizing that $\lambda$ vary for all three configurations, as shown in Fig. \ref{fig5}a. Additionally, $\rho_o$ also varies significantly for configuration C2a. Thus, a different approach is to rewrite Eqns. (\ref{eqn_fn}) and (\ref{eqn_fn2}) as 
\begin{equation}
\frac{ \langle u_s^2\rangle}{ \langle u_l^2\rangle} = f'(n) [1+2\phi(n)g(\phi(n))], 
\label{new_eqn_state}
\end{equation}
where 
\begin{equation}
f(n)' = \frac{(1-n)\rho_o \beta}{(1-\beta \rho_o n)  \left ( 1 + 2 \frac{l_o}{\lambda} \right )}
\end{equation}
and 
\begin{equation}
\phi = \frac{(1-n)\rho_o}{(1-\beta \rho_o n) } \frac{\pi a_{\rm eff}^2}{4}.
\end{equation}
Thus, the objective here is to plot both sides of equation (\ref{new_eqn_state}), using the measured $\langle u_s^2\rangle/\langle u_l^2\rangle$ quantities and also measured values of $\beta$, $\lambda$ and $\rho_o$ for each $n$, instead of their averages. The correctness of the proposed equation of state must then be manifested by a linear one-to-one relation between both plotted quantities of each side of Eqn. (\ref{new_eqn_state}). This is indeed the case as shown in Fig. \ref{fig7}. Here, the solid line corresponds to the equality of both sides of the equation of state. 

In order to obtain $ f'(n) [1+2\phi(n)g(\phi(n))]$ we must also adjust the parameter $\kappa$. This is done by defining for each side of equation (\ref{new_eqn_state}) the measured quantities $Y_i = (\langle u_s^2\rangle/\langle u_l^2\rangle)_i$ and $\tilde{Y}_i = (f'(n) [1+2\phi(n)g(\phi(n))])_i$, where the index $i$ indicates a given measurement for each configuration, that means a given $A$ for C1 and C2b and a given $\phi_o$ for C2a. Therefore, for each configuration we obtain $\kappa$ by minimizing the following objective function
\begin{equation} 
\chi^2 = \sum_{i=1}^{N_m} (Y_i-\tilde{Y}_i)^2, 
\end{equation}
where $N_m$ is the number of measurements for each experimental set of data. In practice, this is done by computing $\chi^2$ for a range of $\kappa$; it has a parabolic shape with a well defined minimum. Following this procedure we obtain $\kappa = 0.468$, $0.577$ and $0.617$ for C1, C2a and C2b, respectively. These values are almost identical to those obtained by doing the adjustment using the global average quantities for $\lambda$, $\beta$ and $\rho_o$.

\subsection{Experimental results: solid cluster's kinetic energy}

The next quantity to compare with experimental measurements is the prediction of the RMS horizontal velocity of particles in the solid phase, given by Eqn. (\ref{ecn_us}). This quantity can be rewritten as 
\beq 
\frac{\sqrt{\langle u_s^2\rangle}}{A\omega} =  C \left [ \frac{\lambda}{A} F(A,\lambda)\right ]^{1/3}, 
\label{ecn_us2}
\eeq 
where 
\beq
C =  \left [ \frac{1-(1-r^2)\delta}{2\pi^{3/2}(1-r^2)} \right ]^{1/3}, 
\eeq
is the only unknown parameter here, and 
\beq
F(A,\lambda) = \sum (v_{\rm coll}/A\omega)^2
\eeq
corresponds to the sum of normalized collisional velocities per period, equal to the normalized dissipated energy presented in Fig. \ref{fig4}c, which we obtain numerically for each $A$ and $\lambda$, being both measured quantities.

\begin{figure}[t!]
\centering
\includegraphics[width=8cm]{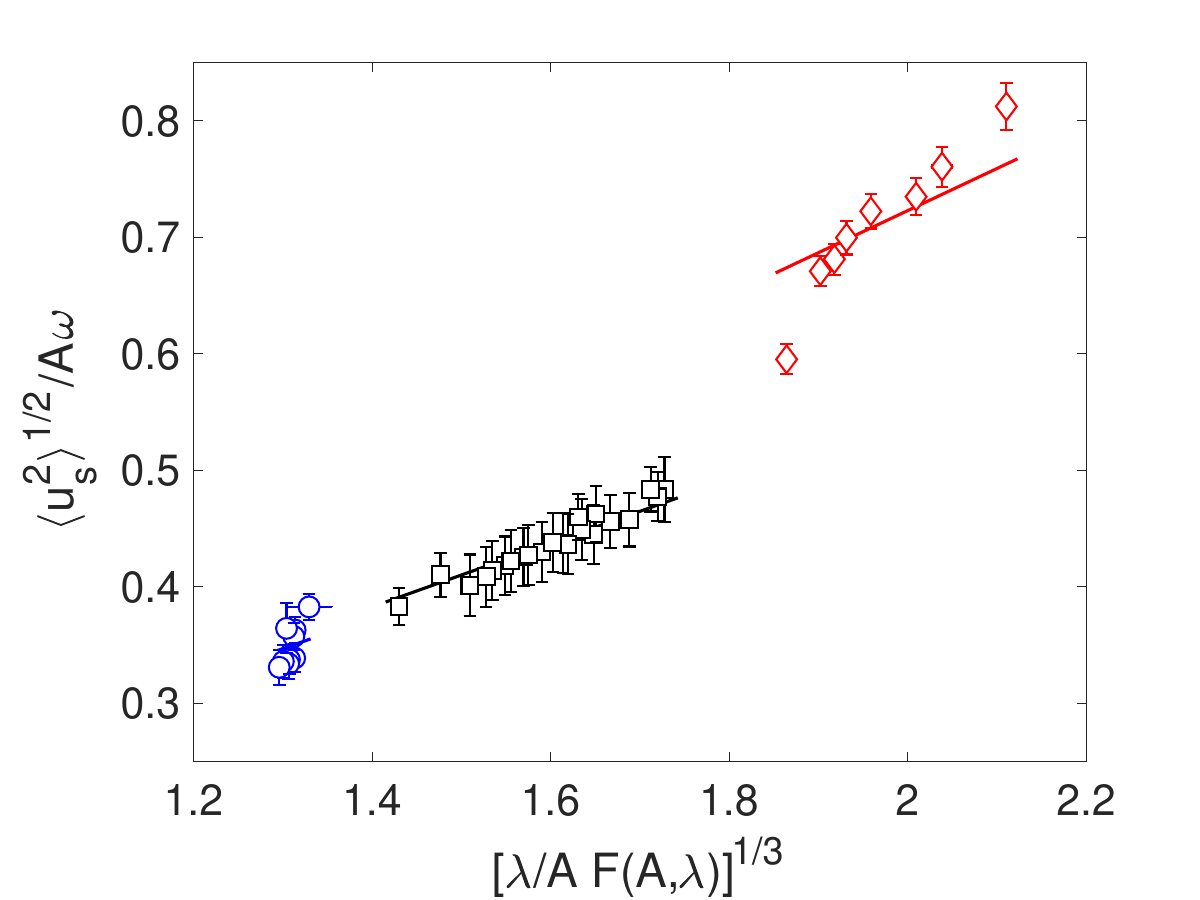}
\caption{Normalized RMS horizontal velocity of particles in the solid phase $\langle u_s^2 \rangle/A\omega$ as function of $\left [ \lambda F(A,\lambda)/A \right ]^{1/3}$. Results for the three configurations are presented: C1 ($\diamond$), C2a ($\circ$) and C2b~($\square$). Solid lines correspond to the model prediction, namely the linear relation $\sqrt{\langle u_s^2\rangle}/A\omega =  C \left [ \lambda F(A,\lambda)/A \right ]^{1/3}$, which is better verified for configuration C2b. The fitted constants for the three configurations are, respectively, $C = 0.36\pm0.02$, $0.27\pm 0.01$ and $0.273\pm0.002$.}
\label{fig8}
\end{figure}

In Fig. \ref{fig8} we present $\langle u_s^2 \rangle/A\omega$ versus $\left [ \lambda F(A,\lambda)/A \right ]^{1/3}$ for the three sets of experimental results. The linear dependence is best observed for configuration C2b, whereas for C1 it works well only in a first approximation as deviations are clearly observed. For C2a, as we only vary $\phi_o$, and not $A$, the data does not vary much around a mean value. For the three cases we have fitted the measurements to the linear relation predicted for these two quantities. The fitted parameter $C$ is $\approx 0.27$ for both C2 data sets, and it is $\approx 0.36$ for C1. Using reasonable values for the restitution coefficient, say $r = 0.8$ and $0.9$, we obtain estimates for the parameter $\delta$: for $r=0.8$, $\delta = 2.3$ and $2.6$ for C1 and C2, respectively; for $r=0.9$, $\delta = 4.7$ and $5.0$ for C1 and C2, respectively. As expected, for higher restitution coefficients, more internal collisions between the two solid phase layers are needed, leading to larger values of $\delta$. Additionally, the numerical values of these estimated values are quite reasonable, considering that $\delta$ takes into account the internal vertical collisions within the bi-layer, for which each particle in one layer can collide with its four neighbors of the other layer.

\subsection{Bifurcation diagram and critical amplitude: effect of frequency, filling density and dissipation}

We now turn to the theoretical prediction of the bifurcation diagram. Keeping all the parameters fixed, we can analyze Eqn. (\ref{eqnGnA}), which gives the solid particle fraction values $n=N_s/N$ that solve this form of the power balance equation for different values of vibration amplitude $A$. This analysis is qualitative and only semiquantitative, as several simplifications have been made. The first is that the mean free path is kept constant, $\lambda = 0.1a$, considering the mode values reported above, ${\rm Mo}(\lambda)\approx 0.08a$ for C1 and ${\rm Mo}(\lambda)\approx 0.12a$ for C2. However, the distribution of $\lambda$ is a function of the forcing amplitude $A$ and the filling density $\rho_o$, and also depends on the forcing frequency $f$. In addition, we fix the anisotropy parameter $\alpha = 2$, a value obtained from measurements performed in a similar way to those presented in Section \ref{subsec_collect_and_collide}, using images obtained with an inclined view of the camera. This quantity is also most likely to be a function of $A$, $f$, and $\rho_o$. Finally, we assume $r = r' = 0.8$ for configuration C2 and $r = r' = 0.92$ for C1, reasonable values of the restitution coefficient for both stainless steel particle collisions and of these with ITO coated glass plates. 

\begin{figure*}[t!]
\centering
\includegraphics[width=8.5cm]{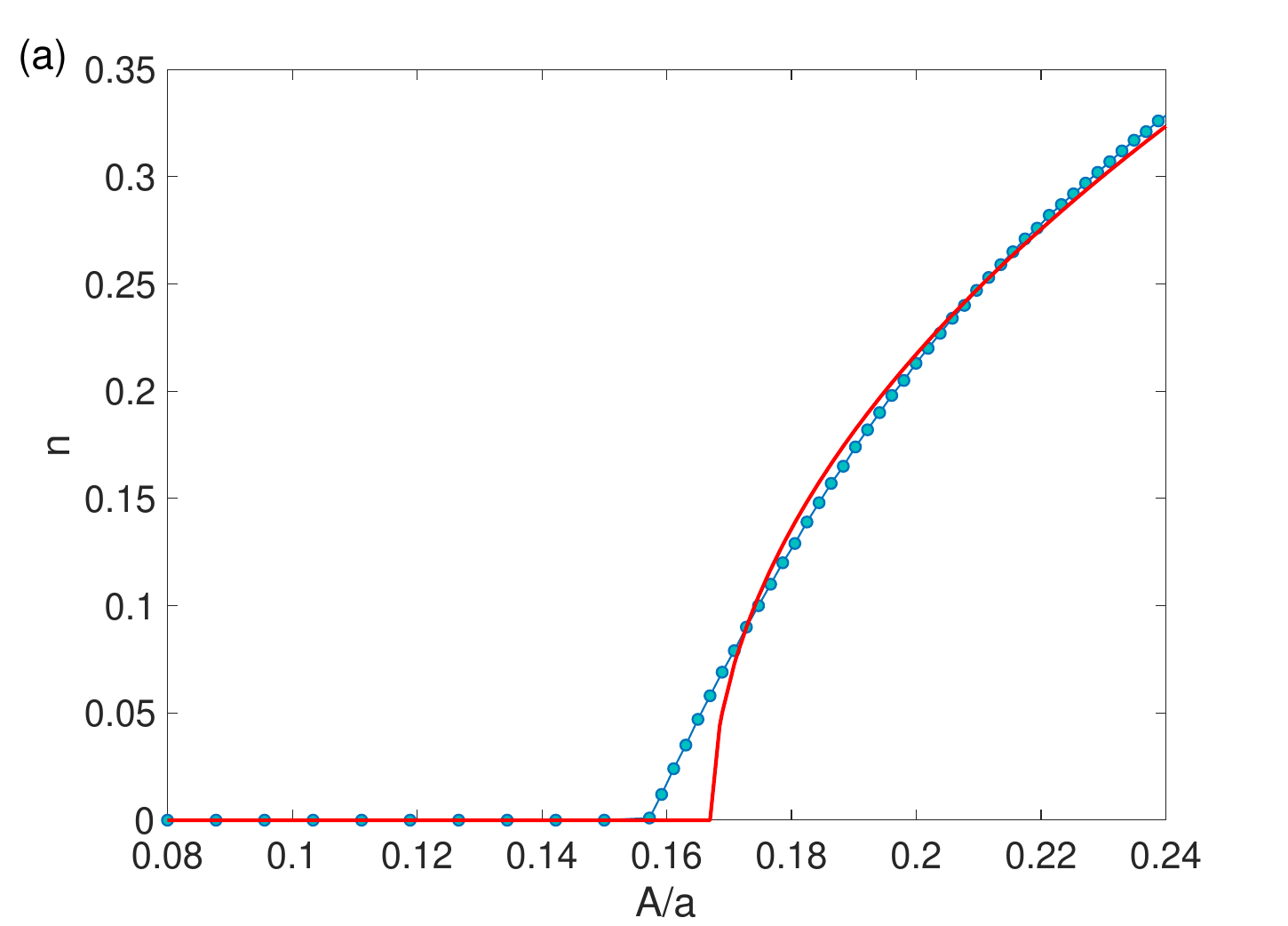}
\includegraphics[width=8.5cm]{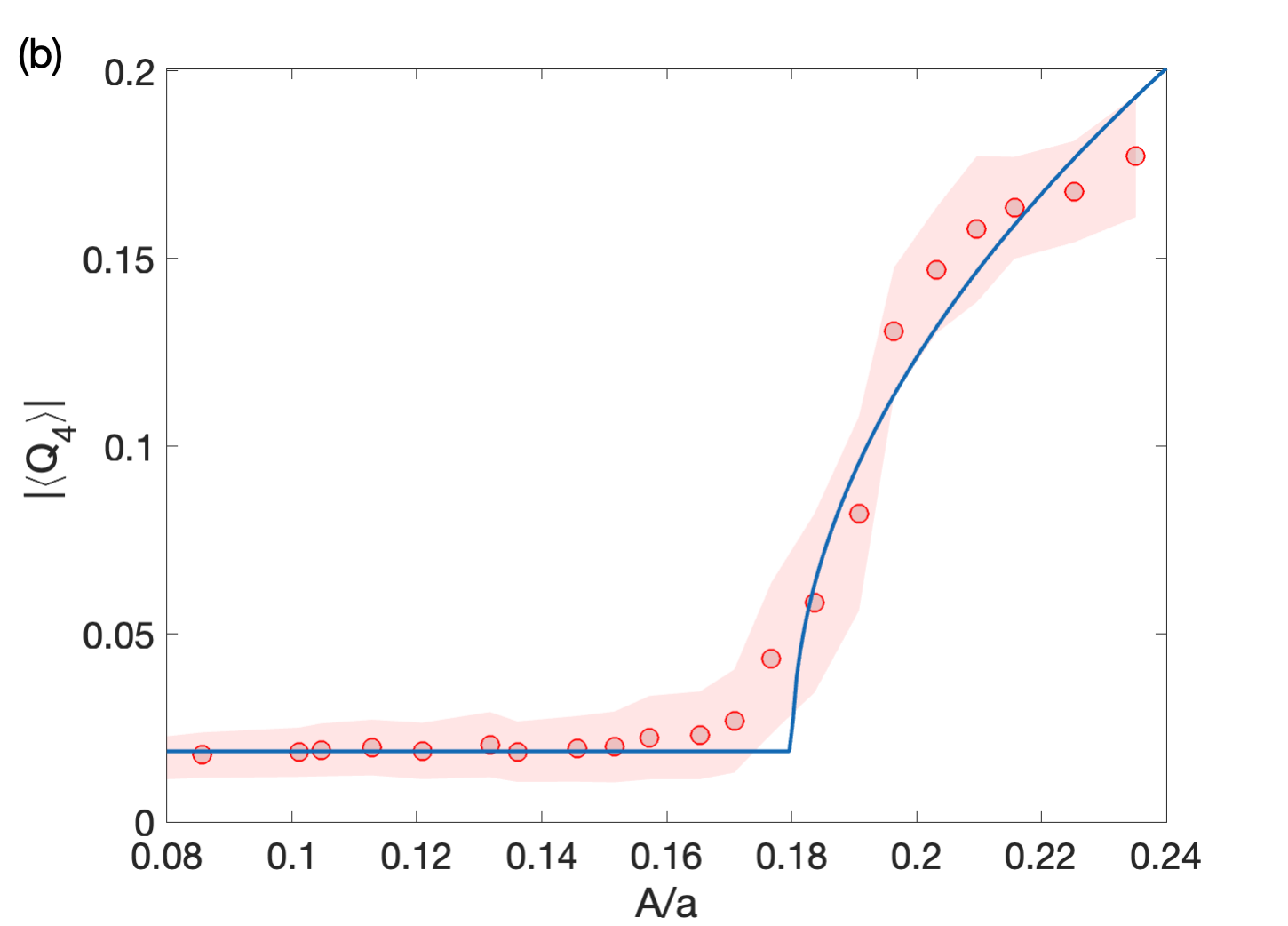}
\includegraphics[width=8.5cm]{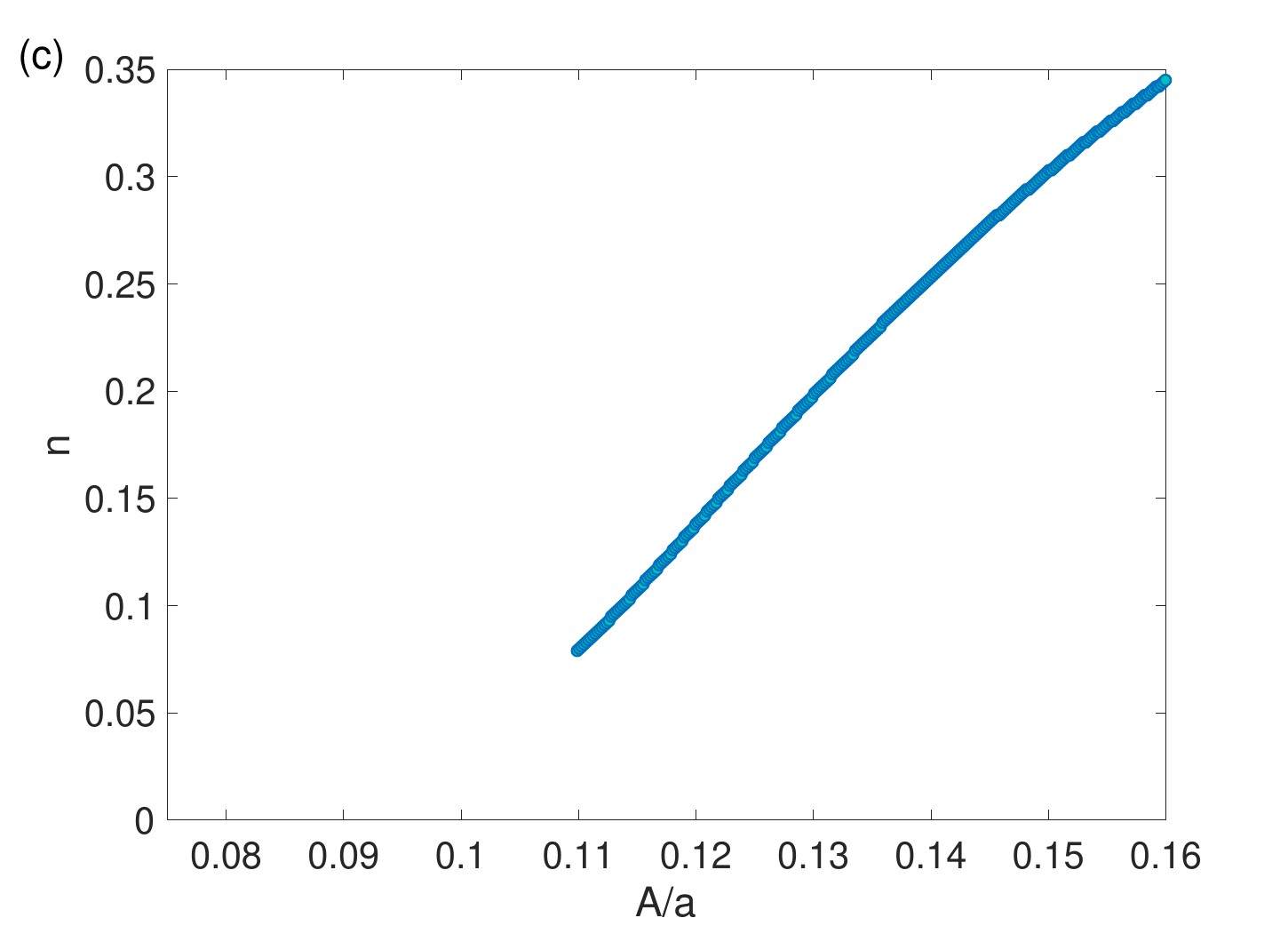}
\includegraphics[width=8.5cm]{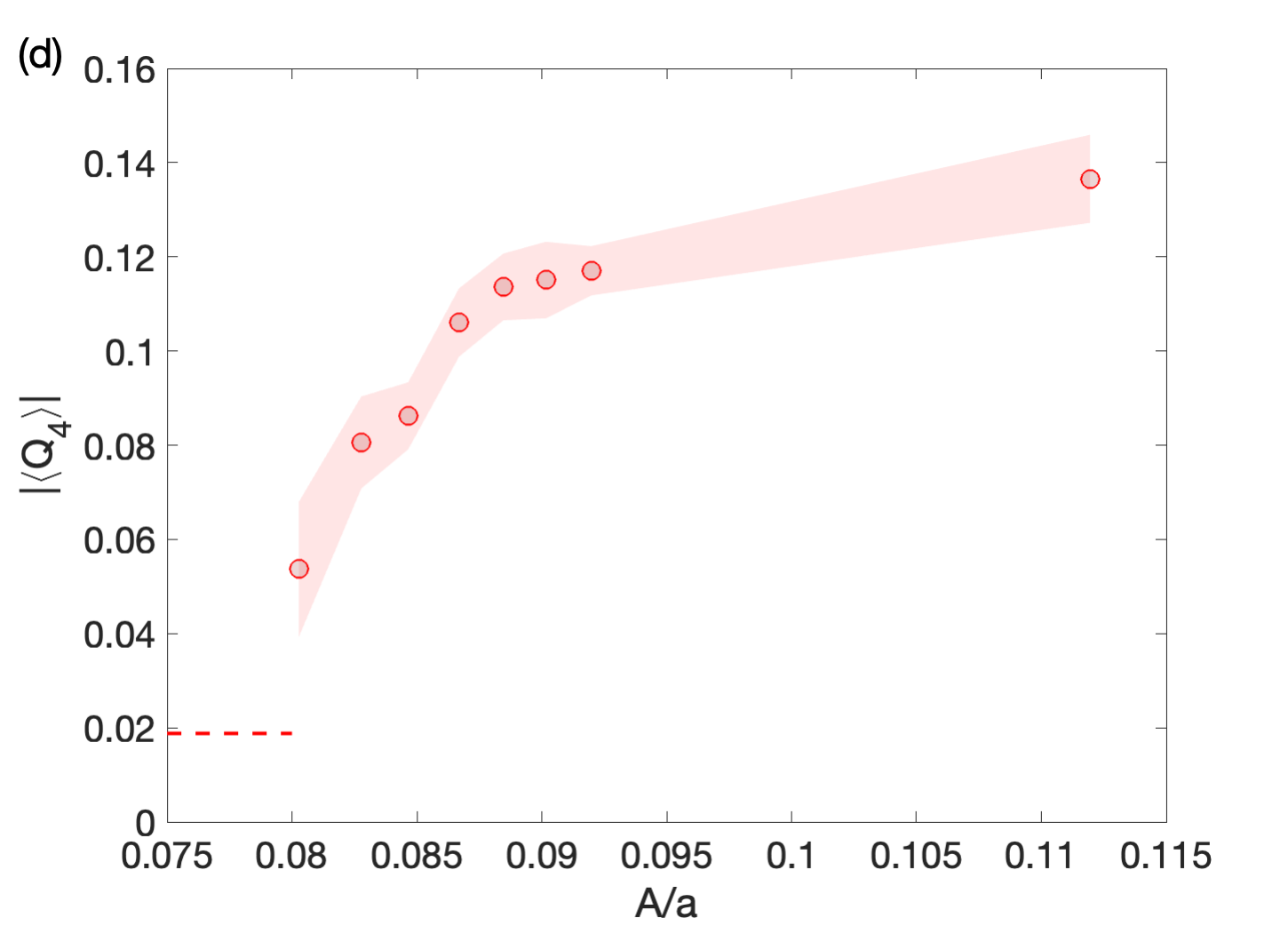}
\caption{Theoretical prediction of the fraction of particles in the solid phase, $n$, versus $A/a$ and comparison with experiments using $|\langle Q_4\rangle|$, for configurations C2 (a,b) and C1 (c,d). (a) Model prediction for configuration C2: $h = 1.94a$, $a = 1$ mm, $f = 80$ Hz, $\rho_o a^2 = 1.2$, $\kappa = 0.6$, $\lambda = 0.1a$, $\alpha = 2$,  $C = 0.27$ and $r = 0.8$. The critical amplitude is $A_c/a = 0.157$. The solid lines show the fits to the supercritical behavior $\sim(A-A_c)^{1/2}$ (details in the main text). (c) Model prediction for configuration C1: $h = 1.83a$, $a = 1$ mm, $f = 80$ Hz, $\rho_o a^2 = 0.976$, $\kappa = 0.47$, $\lambda = 0.1a$, $\alpha = 2$,  $C = 0.27$ and $r = 0.92$. There is a finite jump at $A=0.1099a$, exactly at the maximum dissipation value $A_{\rm max}$ for these parameters (see Fig. \ref{fig4}c). (d)  $|\langle Q_4\rangle|$ versus $A/a$ for configuration C1. The horizontal dashed line shows the small constant value detected in the liquid state, $Q_4^0 = 0.019$.}
\label{fig9}
\end{figure*}

In Figs.~\ref{fig9}a,c we present the prediction of our model for the fraction of solid particles, $n$, as a function of the normalized driving amplitude, $A/a$, for both configurations, C2 and C1, respectively. All other parameters are fixed (values are given in the figure caption). Except for the restitution coefficient, these are all determined by experimental measurements performed with each configuration. We recall that analogously to the zero gravity case \cite{PoschelPRL}, we consider the collect-and-collide regime to be in place for $A\geq A_{\rm max}$, with $A_{\rm max}$ being the amplitude at which the energy dissipation is maximum. Thus, we consider our relevant physical solutions to be in this situation. For configuration C2, the bifurcation is continuous and of second-order type. The transition threshold is $A_c \approx 0.16a$, whereas the measured critical amplitudes are in the range $0.17a-0.20a$ \cite{Castillo2012,Castillo2015}. In fact, panel (b) shows the absolute value of the average bond orientation order parameter, $|\langle Q_4\rangle|$, as a function of $A/a$ for configuration C2b, which is qualitatively very similar to the theoretical prediction presented in panel (a). In fact, we expect $|\langle Q_4\rangle| \sim n$, as most particles in the solid phase have a large value of $|Q_4|$, close to $1$. In both panels, we present fitted curves of supercritical behavior $\sim(A-A_c)^{\frac{1}{2}}$: in (a), $n = \tilde n [(A-{A}_c)/a]^{\frac{1}{2}}$, with ${A}_c/a = 0.167\pm0.001$ and $\tilde n = 1.20\pm0.01$; in (b), $|\langle Q_4\rangle| = Q_4^0 + B [(A-{A}_c)/a]^{\frac{1}{2}}$, with ${A}_c/a = 0.180\pm0.003$, $Q_4^0 = 0.019\pm0.001$ and $B = 0.74\pm0.06$. The quantity $ Q_4^0$ corresponds to a small background value of the fluctuations of the solid cluster in the liquid phase, and has been calculated as the average of $|\langle Q_4\rangle|$ for $A<0.16a$. 

As mentioned above, Fig.~\ref{fig9}c presents the theoretical prediction of $n$ as function of $A/a$ for configuration C1, using $r=r'=0.92$ in this case. The experimental data is shown in panel (d), $|\langle Q_4\rangle|$ versus $A/a$. Both quantities present finite jumps at a critical transition value. For $n$ obtained with our model we obtain $A_c = 0.1099a =A_{\rm max}$, where we remind that the latter is the amplitude corresponding to a maximum dissipation in the collect-and-colloide state. In this case, the transition is given by the onset of the collect-and-collide regime, which overtakes the collective dynamics. We have also mentioned that for configuration C2, the critical amplitudes depend on the dissipation of the system: $A_c = 0.18a$ $(\Gamma_c = 5.1)$ for thin ITO coated glass plates (stronger dissipation) and $A_c = 0.20a$ $(\Gamma_c = 4.6)$ for thick ITO coated glass plates (lower dissipation) \cite{Castillo2015}. However, the transition observed experimentally for configuration C1 is independent of the ITO coating: in both cases, with thin \cite{Castillo2012} and thick coating (this paper), the transition values are the same $A_c = 0.078a$ $(\Gamma_c = 2.0)$. This strengthens the argument that the transition for case C1 is caused by the onset of the collect-and-collide regime, and not by surpassing dissipation.  We attribute the difference in predicted values $A_c$ to the fact that the effective restitution coefficient $r_{\rm eff}=0$ for the solid cluster is an approximation. 

\begin{figure}[t!]
\centering
\includegraphics[width=8cm]{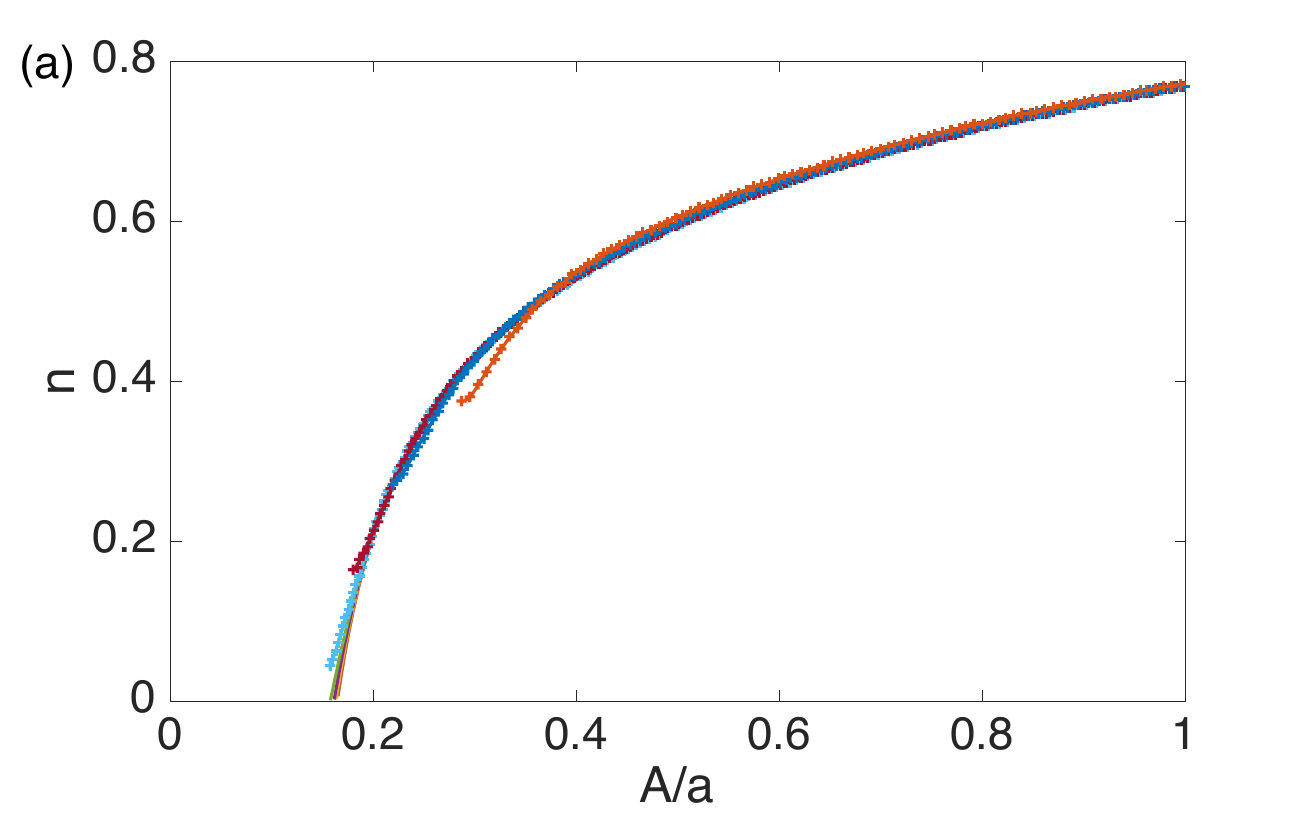}
\includegraphics[width=8cm]{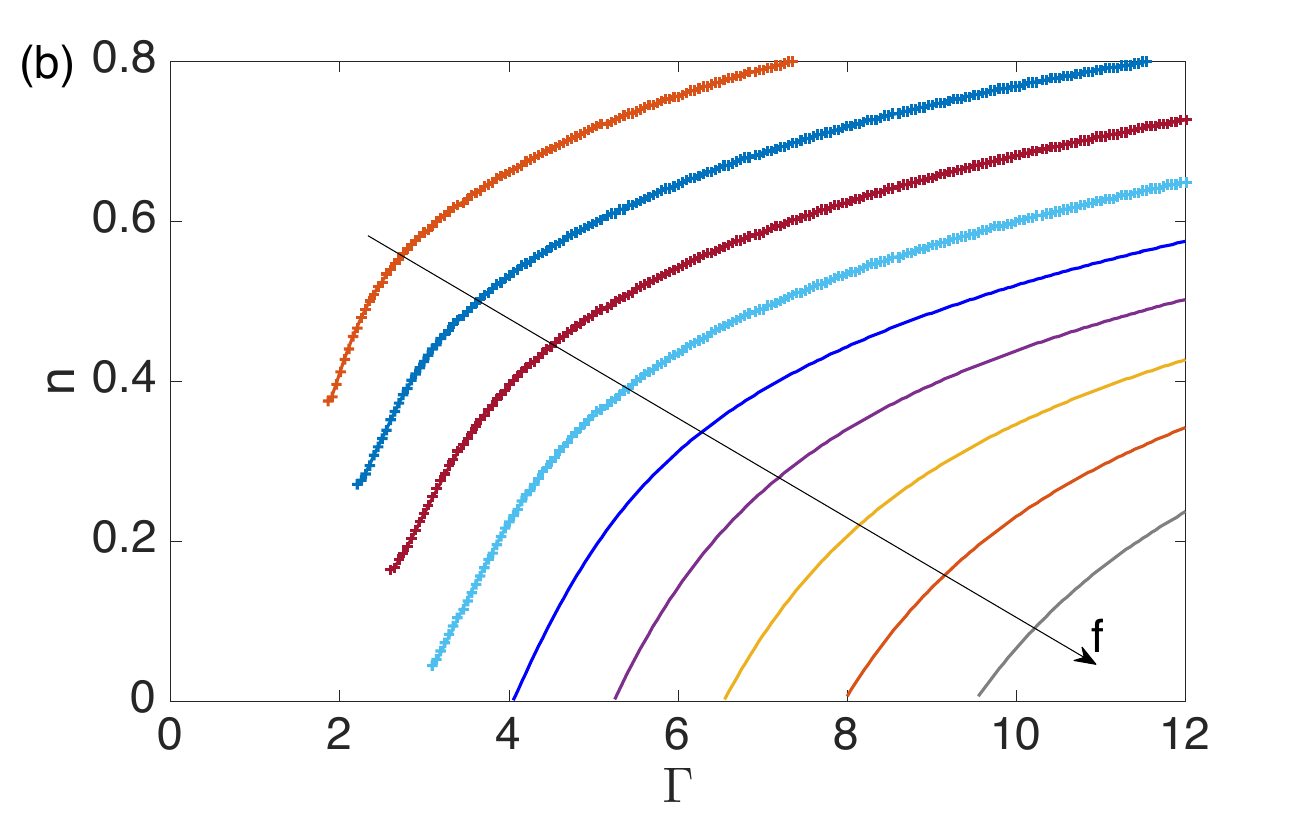}
\caption{Theoretical prediction of the fraction of particles in the solid phase, $n$, versus the normalized driving amplitude $A/a$ (a) and normalized acceleration $\Gamma$ (b) for several frequencies, $f = 40 - 120$ Hz by steps of $10$ Hz ($h = 1.94a$, $a = 1$ mm, $\rho_o a^2 = 1.2$, $\kappa = 0.6$, $\lambda = 0.1a$, $\alpha = 2$,  $C = 0.27$ and $r = 0.8$).}
\label{fig10}
\end{figure}

Keeping all parameters fixed, the transition threshold can change if the frequency is varied. The transition can even change from second to first order type if $f$ decreases below a certain value. This is shown in figure \ref{fig10}, were we plot $n$ versus $A/a$ (a) and $\Gamma$ (b) for several vibration frequencies. In figure (b) the arrow indicates the direction in which $f$ increases, from $40$ Hz to $120$ Hz, by steps of $10$ Hz. Indeed, below $80$ Hz the transition is of first order, diskontinuous. These first order bifurcation curves are shown with symbols $+$, in order to differentiate them from the second order type transitions. Fig. \ref{fig10}a demonstrates that the correct control parameter is the normalized vibration amplitude $A/a$, as all the results collapse on a single curve. It also shows that once the transition is of second order, the critical amplitude remains constant. 

\begin{figure}[t!]
\centering
\includegraphics[width=8cm]{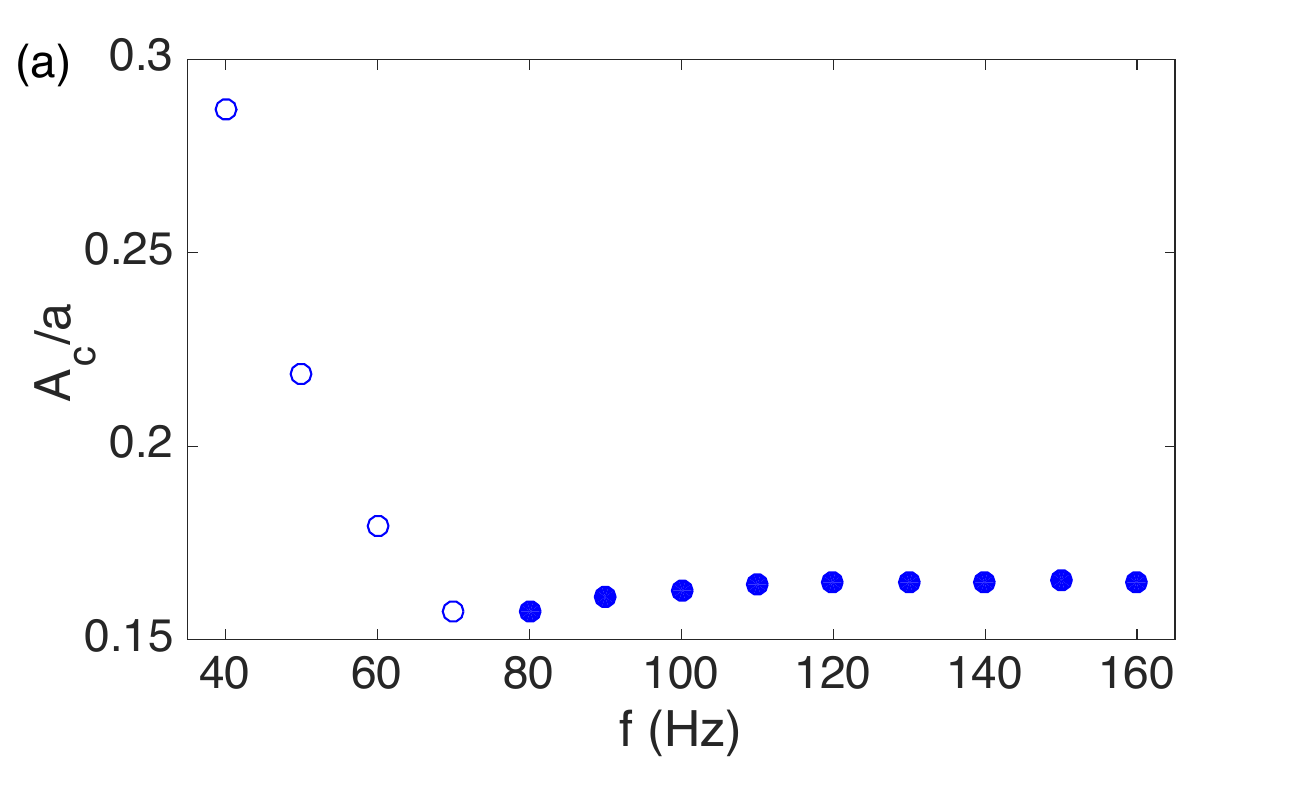}
\includegraphics[width=8cm]{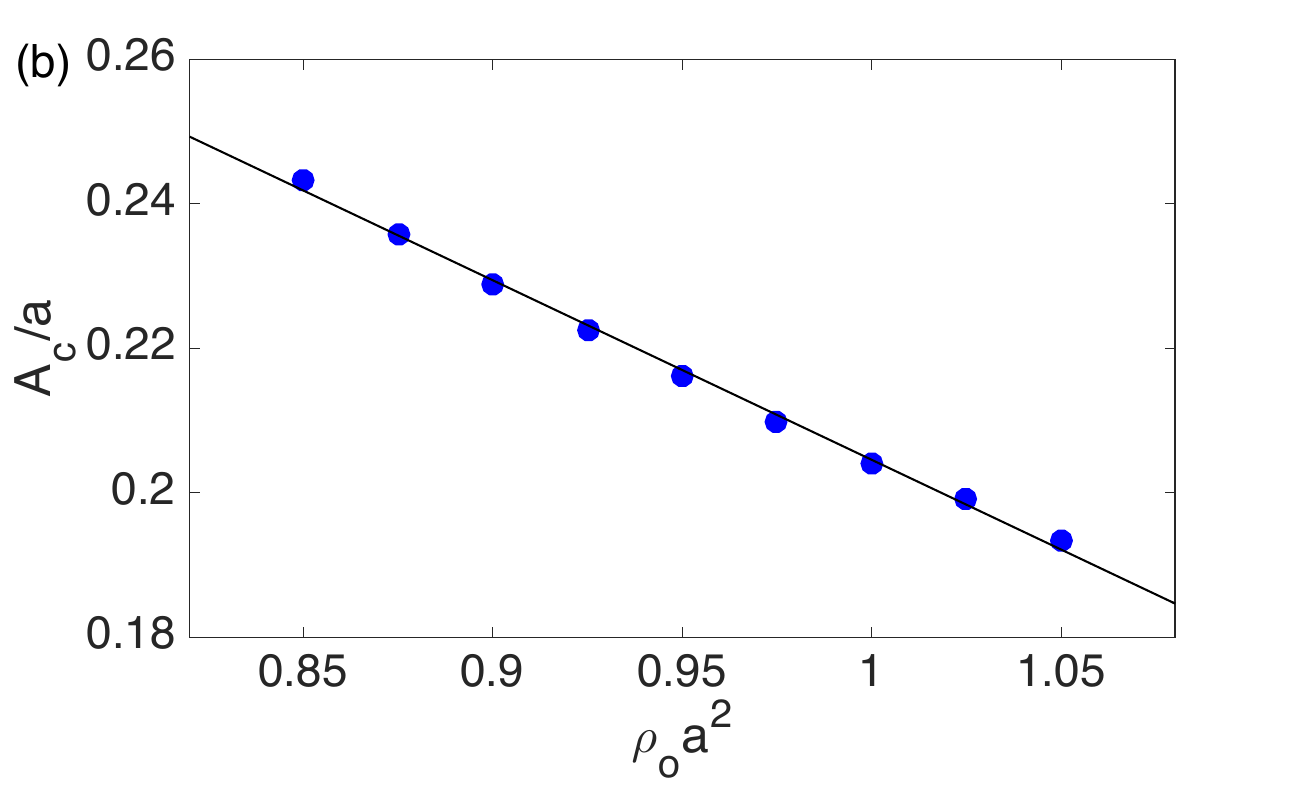}
\includegraphics[width=8cm]{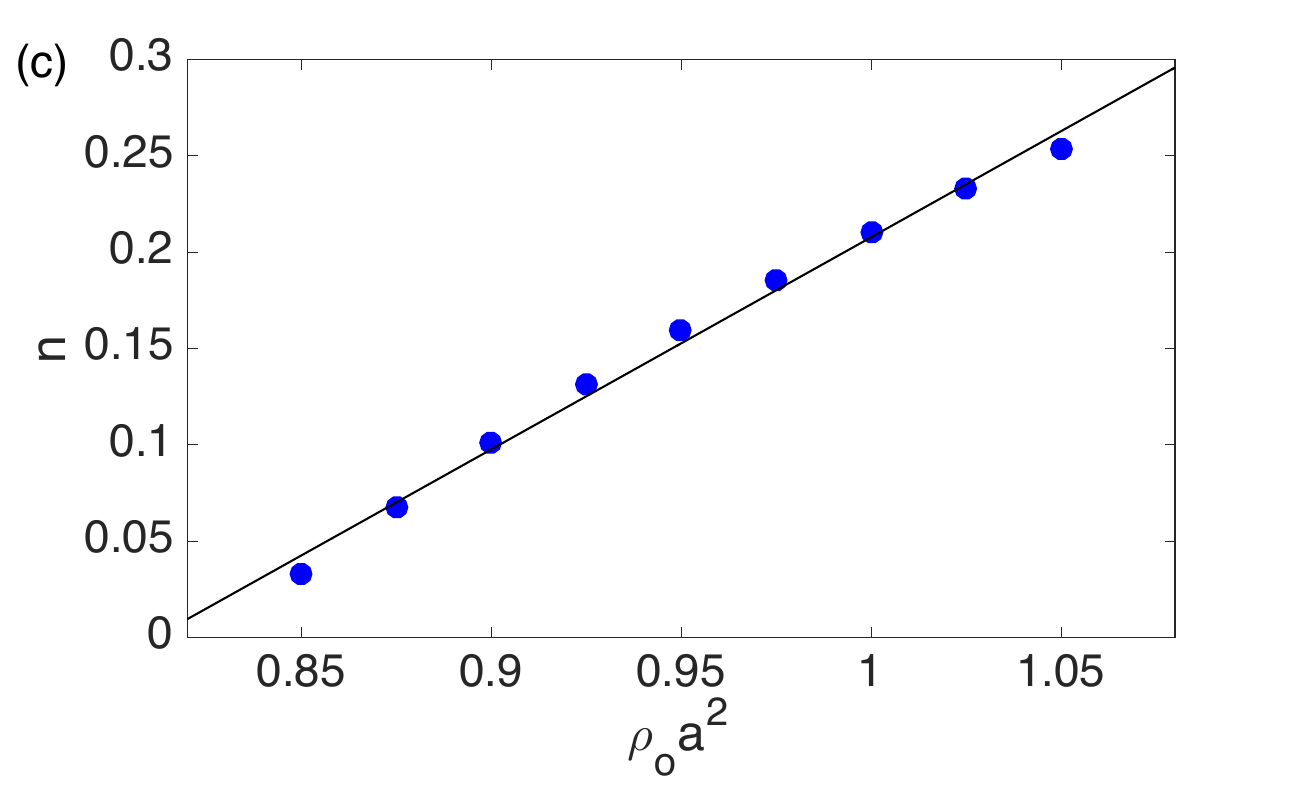}
\caption{(a) Normalized critical amplitude $A_c/a$ as function of vibration frequency $f$. Here, $\rho_o a^2 = 1.2$. Other parameters are defined below. Open (closed) symbols show first (second) order transitions. (b) High frequency critical amplitude $A_c/a$ versus filling fraction $\rho_o$ ($f= 140$~Hz). (c) fraction of particles $n$ in the solid phase as function of filling fraction $\rho_o$ ($f= 140$~Hz and $A = 0.25a$). Solid lines correspond to linear fits, which are shown as guides to the eye. The remaining parameters are $h = 1.94a$, $a = 1$ mm, $\kappa = 0.6$, $\lambda = 0.1a$, $\alpha = 2$,  $C = 0.27$ and $r = 0.8$.}
\label{fig11}
\end{figure}

In Fig. \ref{fig11}a we present the critical vibration amplitude as function of frequency. Open symbols correspond to first order transitions, whereas closed ones to second order type. The threshold first decreases and then it reaches a plateau value at high frequencies. The shape of this curve is very similar to the one obtained by Prevost et. al. \cite{Prevost2004} with molecular dynamic simulations (see Fig. 3a in the cited reference). However, they did not identify first or second order type transitions as we do here, although they did obtain different critical amplitudes (nucleation versus evaporation) depending on the direction the vibration amplitude was varied. We speculate that these numerical measurements found different nucleation and evaporation amplitudes because of the short, finite time measurements. If the simulation time would be made longer, they would have probably found a unique critical amplitude for increasing and decreasing amplitude ramps, as we did in a previously reported experimental study \cite{Castillo2012}.

We can continue to compare our results with Prevost et. al. \cite{Prevost2004}. Fig. \ref{fig11}b shows the high frequency plateau value of the amplitude threshold as function of the filling density $\rho_o$. Here, again all other parameters have been kept fixed with the values used before (Fig. \ref{fig9}). Here, the qualitative comparison with molecular dynamic simulations is quite good. The shape of the transition line between the pure liquid and coexisting liquid and solid phases is quite similar to the result of Prevost et. al., although our results tend to be more linear (see Fig. 3b of \cite{Prevost2004}). Quantitatively, our transition amplitudes are about the double of their molecular dynamic results, but this can be understood because of the lower vertical height used in their simulations ($h = 1.75a$). Finally, to end this comparison, we show in Fig. \ref{fig11}c the fraction of particles in the solid phase, above the transition, as function of the filling density. Again, the qualitative behavior is very similar to the molecular dynamic simulation results of Prevost et al. (see Fig. 3c of \cite{Prevost2004}). In our case however, there is a slight deviation from the pure linear behavior obtained in their work.

\begin{figure}[t!]
\centering
\includegraphics[width=8cm]{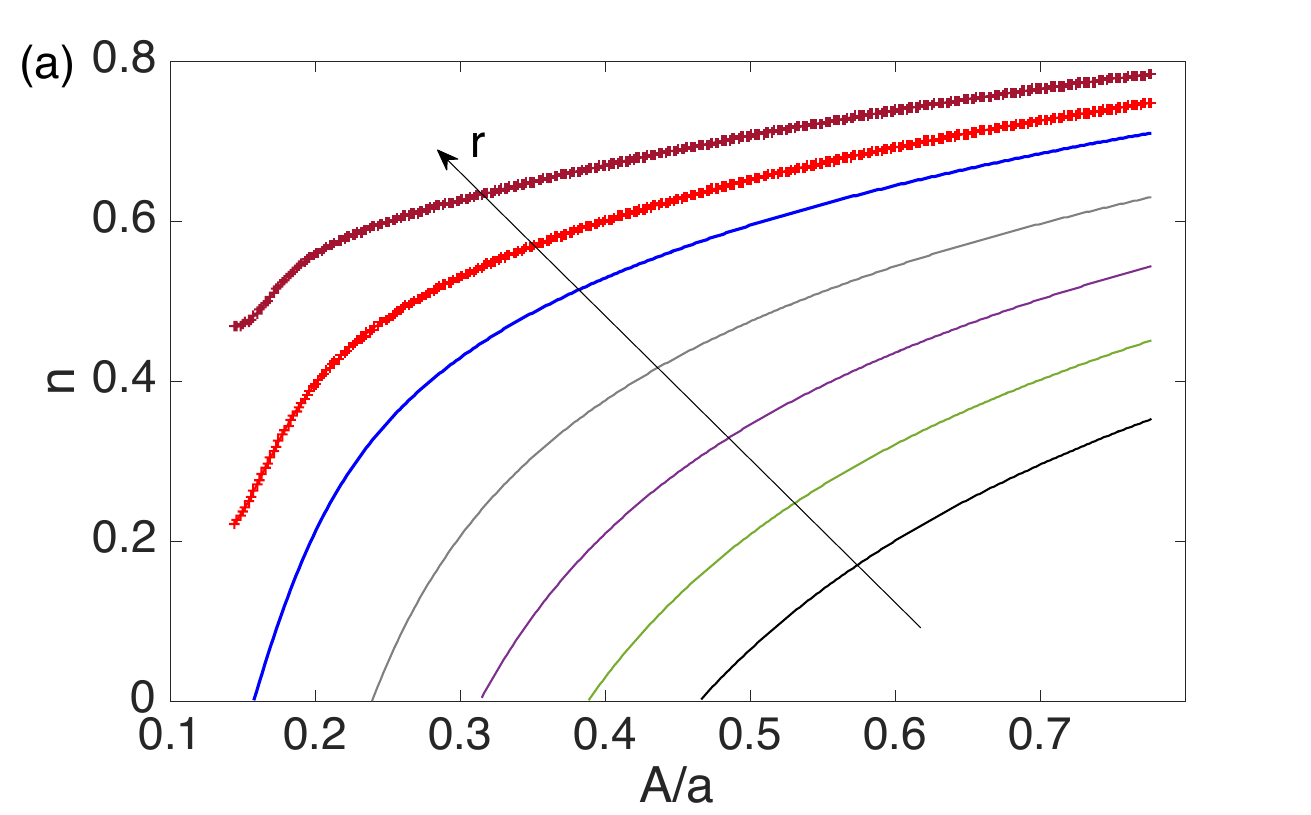}
\includegraphics[width=8cm]{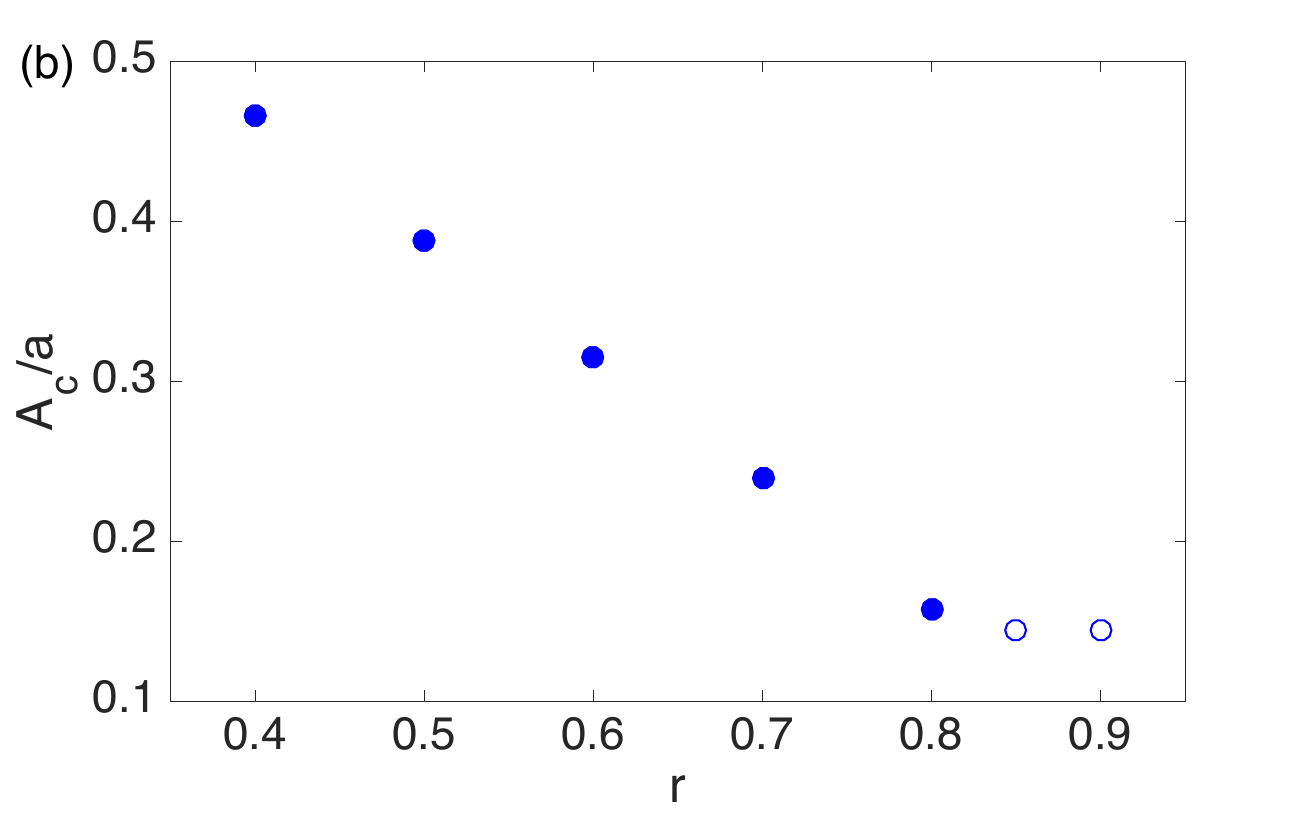}
\caption{(a) Theoretical prediction of the fraction of particles in the solid phase, $n$, versus the normalized driving amplitude, $A/a$, for several restitution coefficients, from right to left $r = 0.4$, $0.5$, $0.6$, $0.7$, $0.8$, $0.85$ and $0.9$. First order type bifurcations are shown with symbols ($+$). (b) Critical amplitude $A_c/a$ versus the restitution coefficient. Open (closed) symbols correspond to first (second) order type transitions. The other parameters are $h = 1.94a$, $a = 1$ mm, $\rho_o a^2 = 1.2$, $\kappa = 0.6$, $\lambda = 0.1a$, $\alpha = 2$ and  $C = 0.27$. }
\label{fig12}
\end{figure}

The transition can also change from second to first order by decreasing the amount of dissipation, as exhibited in Fig. \ref{fig12}a. Fig \ref{fig12}b shows first that when $r$ is varied from $0.4$ to $0.8$, keeping all other parameters fixed, the vibration amplitude threshold decreases linearly. Above $r=0.8$ it is almost constant, and the transition becomes of first order type. The fact that dissipation delays the appearance of the solid-liquid transition is consistent with experimental and molecular dynamic simulation results reported by Vega Reyes and Urbach~\cite{Vega2008}.  

Finally, Lobkovsky et. al. performed molecular dynamic simulations in order to study the effect of forcing and dissipation on this phase transition \cite{Lobkovsky2009}. In particular, they found that energy injection in the vibrated system is quite different in the liquid and solid phases. They claim that this suggests the existence of an effective surface tension, a quantity that, in spite of being quite low, has been shown to be measurable in this system \cite{Luu2013}. Both the large difference in injected energies and the existence of a surface tension were not present in simulations done with a random forcing, which is consistent with the importance of the solid cluster vertical dynamic synchronization for phase separation. In the case of the vibrated system, they found a difference of about a factor of 1/2, being the injected energy larger in the liquid phase. Using the definitions of our model, for $A/a$ in the range $0.2-0.3$, we obtain $I_s^*/\langle I_l \rangle \sim 0.2-0.8$ depending on the restitution coefficient and the type of transition, first or second order.

Finally, from Eqn. (\ref{ecn_us2}) we can estimate the surface tension of the solid cluster, $\gamma = \frac{m}{2} \langle u_s^2\rangle/a$ \cite{Luu2013}. For configuration C2, its measured value is $\gamma = 2.9$ $\mu$N for $\Gamma =6.3$, well above the critical acceleration $\Gamma_c = 5.1$ \cite{Luu2013}. Our model allows to obtain $\gamma$ as function of the forcing amplitude. For configuration C2 and $\lambda =0.1a$ we obtain $\gamma = 2.3$ $\mu$N for $\Gamma =6.3$, in rather in good agreement with the experimental value. We recall that our pressure balance equation neglects surface tension effects; this is demonstrated a posteriori by the validation of this assumption through Eqn. (\ref{new_eqn_state}), as shown in Fig. \ref{fig7}.

\section{Conclusions}

In this paper we propose a phenomenological model that explains the microscopic origin of the solid-liquid phase transition. We also compare our theoretical predictions with experiments from three different sets of data, that is from three different experimental configurations. Comparisons are also done with experimental and molecular dynamic simulation results from other groups. 

Based on the experimental observation of the synchronization of the dynamics of the solid cluster with the vibration of the cell, we propose to model the bi-layer crystalline structure as a completely inelastic particle with internal degrees of freedom, namely the mean free path $\lambda$ and horizontal 'granular temperature'', quantified by~$\langle u_s^2 \rangle$. Similarly to Sack et al. \cite{PoschelPRL}, but including gravity, its vertical dynamics is then obtained by the solution of an effective particle of mass $N_s m$ with zero restitution coefficient $r_{\rm eff} = 0$ confined between two vibrating plates with a separation $h'$, which is given by the bi-layered square structure with a mean free path $\lambda$. 

Following Luding~\cite{Luding2009}, we write down the general form of the equation of state for a granular fluid in 2D. Here, our quasi-2D system is projected to an effective two-dimensional one, by allowing the diameter of the disk to have an effective value, parametrized by the factor $\kappa$. For the liquid phase, the collisional term of pressure 2$\phi g(\phi)$ is modeled by an expression for the pair correlation function that has been shown to be valid up to moderate densities ($g(\phi) \equiv g_2(\phi) $ \cite{Luding2009}). For the solid phase, as such model is not well established for large densities, even less so for a quasi 2D bi-layered structure,  in order to compute the collisional term we use the measurement of the mean free path instead ($\phi g(\phi) = l_o/\lambda$). The adjustment of the equation of state to the experimentally measured quantities is very satisfactory, allowing good fits of the effective diameter factor $\kappa$ for the three experimental configurations.

 By computing the real dissipated power in the solid cluster, $\langle D_s \rangle$, and imposing its equality with the dissipated power of the effective inelastic particle, $D_s^*$, we obtain an expression for the solid cluster's RMS velocity $\langle u_s^2 \rangle^{1/2}$.  To do so, the internal vertical collisions have to be taken into account through the collisional frequency and the collision factor $\delta$. Through the fit of experimentally measured $\langle u_s^2 \rangle^{1/2}$ as function of $A$ and $\lambda$, estimated values of $\delta$ are deduced, with quite reasonable values. 
 
The last parameter of the model is the anisotropy parameter $\alpha$, defined by the ratio of vertical and horizontal RMS velocities in the liquid phase: $\sqrt{\langle v_l^2 \rangle} = \alpha \sqrt{\langle u_l^2 \rangle}$. Although this parameter is most likely a function of filling density, vibration amplitude and frequency, we have used a fixed value $\alpha = 2$ obtained from measurements of particle vertical and horizontal dynamics using images obtained with an inclined view of the experimental setup (measured values vary between $1.5$ and $1.8$). 

We finally obtain a simplified form of the power balance equation, $G(n,A) = 0$, given by Eqn. (\ref{eqnGnA}). For a given set of system and forcing parameters, this allows to obtain the solutions of $n$ for each $A$. In order to study the physically relevant solutions, we constrain the vibrations amplitudes to be equal or greater than the value at which the dissipated energy is maximized, that is, for $A\geq A_{\rm max}$. The obtained bifurcation diagram fits well both qualitatively and semi-quantitatively the experimental observations. The model predicts the existence of both first and second order transitions (abrupt versus continuous). The transition can be changed from second to first order by varying the forcing frequency or the system's dissipation. The amplitude threshold is indeed delayed for more dissipative collisions, in agreement with experimental and molecular dynamic results.

\acknowledgments
This work was supported by FONDECYT grants 1150393 and 1221597.

\end{document}